\documentclass[11pt,oneside,reqno]{amsart}
\usepackage[T1]{fontenc}
\usepackage[latin9]{inputenc}
\usepackage[a4paper]{geometry}
\usepackage{color}
\usepackage{amstext}
\usepackage{amsthm}
\usepackage{amssymb}
\usepackage{setspace}
\usepackage[authoryear]{natbib}
\usepackage{graphicx}

\usepackage{ulem}

\makeatletter

\numberwithin{equation}{section}
\numberwithin{figure}{section}
\theoremstyle{plain}
\newtheorem{thm}{\protect\theoremname}
\theoremstyle{remark}

\theoremstyle{plain}
\newtheorem{lem}[thm]{\protect\lemmaname}

\usepackage{amsfonts}
\usepackage{bbm}
\usepackage{color}

\newtheorem{cor}{Corollary}\newtheorem*{asm}{Assumption}\theoremstyle{definition}

\usepackage[rightcaption]{sidecap}
\usepackage{comment}

\providecommand{\remarkname}{Remark}

\makeatother

\providecommand{\lemmaname}{Lemma}
\providecommand{\remarkname}{Remark}
\providecommand{\theoremname}{Theorem}

\begin{document}
\title{Regression adjustment in completely randomized experiments with many covariates}
\author{Harold D. Chiang}
\address{Department of Economics, University of Wisconsin-Madison, 1180 Observatory Drive Madison, WI 53706-1393, USA.}
\email{hdchiang@wisc.edu}
\author{Yukitoshi Matsushita}
\address{Graduate School of Economics, Hitotsubashi University, 2-1 Naka, Kunitachi, Tokyo 186-8601, Japan.}
\email{matsushita.y@r.hit-u.ac.jp}
\author{Taisuke Otsu}
\address{Department of Economics, London School of Economics, Houghton Street, London, WC2A 2AE, UK.}
\email{t.otsu@lse.ac.uk}
\thanks{We thank Haoge Chang, Xinwei Ma, Ruonan Xu, and the participants at the Conference on Estimation and Inference on Econometric Models 2022 in Toulouse and Econometrics Seminar at University of California San Diego for their valuable comments and discussions. We thank Haonan Miao for his excellent RA works. This research was partly supported by the JSPS KAKENHI grant number 18K01541 (Matsushita). All remaining errors are ours.}
\begin{abstract}
This paper investigates estimation and inference for average treatment effects in completely randomized experiments when researchers observe potentially many covariates. {  Within Neyman's (1923) design-based framework, allowing the number of covariates to grow more slowly than the sample size, we demonstrate that a cross-fitted regression adjustment estimator--adapted from \citet{aronow-middleton2013}--exhibits more favorable asymptotic properties than existing alternatives, such as Lin's (2013) regression adjustment estimator and the bias-corrected estimator of \citet{lei2021regression}.} For inference, we derive the first- and second-order terms in the stochastic expansions of regression-adjusted estimators, analyze the higher-order behavior of existing inference procedures, and introduce a modified  version of the HC3 standard error. The proposed methods extend naturally to stratified experiments with large strata. Simulation studies show that the cross-fitted estimator, in combination with the modified HC3, provides accurate point estimates and reliable size control across a wide range of data-generating processes.


\end{abstract}

\maketitle

\section{Introduction}

Randomized controlled trials (RCTs) remain among the most fundamental and influential tools for causal inference, widely employed by empirical researchers across the natural, social, and biomedical sciences. See, for instance, \citet*{fisher1925statistical,fisher1935}, \citet*{neyman1923application}, and \citet*{kempthorne1952design} for foundational developments, and \citet*{imbens2015causal} and \citet*{rosenberger2015randomization} for modern textbook treatments. Statistical inference in RCTs is typically approached from one of two distinct perspectives: the finite population and the superpopulation frameworks. First introduced by \citet*{neyman1923application}, the finite population perspective treats potential outcomes as fixed, with randomness arising solely from the treatment assignment mechanism. In contrast, the superpopulation approach assumes that observed units are independently drawn from a hypothetical infinite population. While both perspectives are influential and widely adopted, theoretical developments under the finite population framework---especially in more complex settings---remain relatively underexplored. This paper adopts the finite population perspective and examines causal inference using regression adjustment methods for RCTs.\footnote{It is not our intention to advocate for either perspective; see \citet*{reichardt1999justifying} for a philosophical comparison of the two.}

In various RCTs, researchers usually collect covariates that are predetermined characteristics of the experimental subjects and conduct regression adjustments to estimate treatment effects of interest since regression adjustments can potentially reduce variability of the estimates (see, for example, Section 7 in Imbens and Rubin, 2015). However, different opinions exist on whether to adjust for covariates; in an influential work, \citet*{freedman2008regressiona, freedman2008regression} criticized the practice of using regression adjustment for RCTs with three critiques: (i) lack of efficiency guarantee of ad hoc regression adjustment over the unadjusted estimator, (ii) inconsistency of the classical regression variance estimator, and (iii) presence of a bias term of order $O_{p}(n^{-1})$. When the number of covariates is treated as fixed, the first two critiques have been addressed by \citet*{lin2013agnostic}, in which the author suggested running a regression of the observed outcomes on the treatment variable, covariates, and their interactions. {  \citet*{lin2013agnostic} showed that (i) this regression adjusted estimator is consistent and asymptotically more efficient than the simple difference in means estimator without regression adjustment, and (ii) this efficiency improvement is ensured despite arbitrary misspecification of the conditional mean function (called the no-harm property; see also \citet*{negi2021} for analogous results in the superpopulation framework).} In addition, \citet*{lin2013agnostic} showed that the heteroskedasticity robust variance estimators for linear regression is asymptotically conservative and thus provides valid size control. Recently, \citet*{chang2021exact} address the remaining criticism by providing analytic exact bias correction formulae for the regression adjustment estimators in \citet*{freedman2008regressiona, freedman2008regression} and \citet*{lin2013agnostic}. Thus far, at least under the asymptotic framework where the number of covariates held fixed, Freedman's critiques on regression adjustment for RCTs have been addressed.

In addition to these remarkable progresses, attempts have been made to study asymptotic regimes that allow the number of covariates to grow with the population size. Such analyses are empirically important because in many RCT studies, researchers record a sizable set of covariates whose dimension is often not negligible compared to the number of experimental subjects. Indeed, in such scenarios, theoretical guarantees derived under fixed dimensionality may be far less than compelling; with a sizable number of covariates, the bias, oftentimes non-negligible, becomes even more problematic. In such asymptotic environments, an important recent contribution came from \citet*{lei2021regression}; under fairly mild conditions, they establish asymptotic normality permitting growing number of covariates, and characterize the leading term of the bias for the regression adjustment estimator of \citet*{lin2013agnostic}. They go one step further by providing an analytic bias-correction estimator. Despite its promising theoretical guarantees, their proposed bias-corrected estimator does not appear to be nearly bias free in their simulation studies when the DGPs contain more nonlinearity as well as larger numbers of covariates. As a practical solution, they further recommend a trimming procedure for covariates to get around the unreliable finite sample bias performances of their bias-corrected estimator. Nevertheless, the means to effectively tackle the bias problem without resorting to artificial modification of the covariates remain unclear. 

In this paper, we contribute to the endeavor of understanding regression adjustment in multiple fronts. {  First, we study higher-order properties of Lin's (2013) regression adjustment estimator, Lei and Ding's (2021) bias-corrected estimator, and the cross-fitted estimator proposed by \citet{aronow-middleton2013} when the number of covariates grows at a slower rate than the sample size under completely randomized experiments, and show that the cross-fitted estimator possesses improved asymptotic properties compared to the others.} Second, we derive a finer asymptotic variance expression for the estimators that takes into account of the higher-order term. As pointed out in Lei and Ding (2021, Section 4.3), the asymptotic variance of the regression adjustment estimators can deviate significantly from the theoretical ones in finite samples, especially when the dimensionality and/or nonlinearity in the DGPs is non-negligible. This further motivates us to propose an alternative bias-corrected version of the HC3 standard error. The simulation studies unveil supporting evidences that the cross-fitted estimator has favorable performances robustly over a variety of scenarios. Coupled with our bias-corrected HC3, it delivers more precise inference results than existing alternative estimation and inference methods when researchers utilize a modest or large number of covariates for causal inference in RCTs. Our methodology is also extended to cover stratified experiments with large strata.


In both social and natural sciences, researchers often find RCTs involve a sizable number of available covariates in their empirical applications. To formally cope with such settings, \citet*{bloniarz2016lasso} and \citet*{wager2016high} studied regression adjustments by machine learning techniques in a high-dimensional setup where the dimensionality $p$ may be larger than the population size $n$. On the other hand, \citet*{lei2021regression} investigated the situation where $p\ll n$ but $p$ may grow with $n$, and developed a bias correction method for the regression adjustment estimator; as eloquently reasoned by \citet*{lei2021regression}, this moderately growing $p$ asymptotics is of particular importance in a wide range of applications that involve RCTs. {  This paper employs the same setup as \citet*{lei2021regression} and focuses on the case of $p\ll n$.}

\subsection{Relationship to the literature}

This paper is built upon a growing body of the important recent forays into innovating theory of RCTs under finite population asymptotics; these include but are not limited to, \citet*{freedman2008regressiona, freedman2008regression}, \citet*{lin2013agnostic}, \citet*{tan2014second}, \citet*{aronow2014sharp}, \citet*{dasgupta2015causal}, \citet*{bloniarz2016lasso}, \citet*{wager2016high}, \citet*{fogarty2018regression}, \citet*{li2018asymptotic}, \citet*{abadie2020sampling}, \citet*{li2020rerandomization}, \citet*{chang2021exact}, \citet*{imbens2021causal}, and \citet*{lei2021regression}. In particular, \citet*{tan2014second} studies the higher-order asymptotics for regression in various design-based setups assuming fixed covariate dimensionality. {  In a recent independent work, \citet*{lu2025} proposed an alternative bias-corrected estimator and studied asymptotic properties when $p\ll n$ and $p\sim n$. For the case of $p\ll n$ considered in this paper, we investigate asymptotic properties of different estimators up to second-order terms.} It is also closely related to the studies of regression models with many regressors under superpopulation setups such as, e.g. \citet*{cattaneo2018alternative,cattaneo2018inference,cattaneo2019two}, to list a few. The idea of cross-fitting or sample splitting has been widely applied in causal inference literature; in fact, it is a common strategy to reduce bias terms in many semiparametric and high-dimensional models, see, e.g., \citet*{schick1986asymptotically}, \citet*{zheng2011cross}, \citet*{chernozhukov2018double}, \citet*{newey2018cross}, \citet*{spiess2018optimal}, \citet*{bradic2019sparsity}, to list a few.


{  The idea of cross-fitting is well established in RCT contexts. For instance, \citet{aronow-middleton2013} and \citet{wu2018loop} discuss unbiased estimation of the average treatment effect using robust moment conditions and leave-one-out procedures (see also \citet{williams1961} and \citet{robins1999} for related antecedents). In contrast to \citet{aronow-middleton2013}, who derive unbiased estimators under independent treatment assignments, this paper treats the number of treated units as deterministic and accommodates potential dependence among treatment assignments. By incorporating the treatment-covariate interactions highlighted by \citet{lin2013agnostic}, the cross-fitted estimator considered in this paper not only retains the low-bias advantages of cross-fitting but also performs robustly in settings with a large number of covariates.}

Our work sheds new light on these literatures by providing a bias-corrected estimation procedure that combines the idea of cross-fitting and efficient regression-assisted estimation for RCTs, and further establishes formal theoretical justification for its advantages in performances for models in RCTs with large numbers of covariates under design-based finite population asymptotics.\footnote{  In a recent working paper by \citet{matsushita2025}, two of the authors of this paper studied analogous issues under the sampling-based superpopulation setup, and developed an optimal covariates selection criterion and higher-order accurate standard error.}

\section{Methodology\label{sec:setup}}

Consider a treatment-control RCT, where $y_{i}(1)$ and $y_{i}(0)$ are potential outcomes of unit $i=1,\ldots,n$ for treatment and control, respectively, and $T_{i}$ is an indicator for assignment ($T_{i}=1$ corresponds to the treatment, and $T_{i}=0$ corresponds to the control). {  This paper focuses on the completely randomized experiment, where a researcher randomly assigns $n_1$ units to the treatment group and $n_0=n-n_1$ units to the control group (see, Chapter 4.4 of Imbens and Rubin, 2015). More precisely, the experimenter deterministically chooses $n_1$ and $n_0$, and treatment units are randomly drawn size-$n_1$ subset of $\{1,\ldots,n\}$ uniformly over all possible $n!/(n_1!n_0!)$ subsets. We note that this setup is commonly applied in the RCT literature using the design-based analysis (e.g., \citet{freedman2008regressiona}, \citet{lin2013agnostic}, and \citet{lei2021regression} discussed above), and that the treatment variables $(T_1,\ldots,T_n)$ are dependent in contrast to the iid sampling.}\footnote{  \citet{aronow-middleton2013} adopted the iid sampling to study unbiasedness of their estimation method that covers the cross-fitted estimator defined below. The focus of this paper is the higher-order asymptotic property of the cross-fitted estimator with the growing number of covariates under the completely randomized experiment.}

We wish to conduct estimation and inference on the average treatment effect
\[
\tau=\mu_{1}-\mu_{0},\quad \mbox{where }\mu_{t} = \frac{1}{n}\sum_{i=1}^{n}y_{i}(t)\mbox{ for }t=0,1,
\]
based on the observed outcome
\[
Y_{i}=y_{i}(T_{i}),
\]
and $(p-1)$-dimensional pretreatment covariates $x_{i}$. In this paper, we employ the finite population perspective \citep{neyman1923application}, where the potential outcomes $y_{i}(1)$ and $y_{i}(0)$ are non-random and randomness comes solely from the treatment indicator $T_{i}$ (see, e.g., \citet*{imbens2015causal}, for an overview).

The simplest estimator of $\tau$ is the difference in means
\[
\hat{\tau}^{\mathrm{dif}}=\frac{1}{n_{1}}\sum_{i=1}^{n}T_{i}Y_{i} - \frac{1}{n_{0}}\sum_{i=1}^{n}(1-T_{i})Y_{i},
\]
where $n_{1}$ and $n_{0}$ are the sizes of the treatment and control groups, respectively. Although this estimator is unbiased and asymptotically normal, \citet*{lin2013agnostic} showed that a regression adjustment using $x_{i}$ yields a more efficient estimator than $\hat{\tau}^{\mathrm{dif}}$. This regression adjustment estimator $\hat{\tau}^{\mathrm{adj}}$ is obtained as the OLS coefficient on $T_{i}$ from the regression of $Y_{i}$ on $(1,T_{i},(x_{i}-\bar{x})^{\prime},T_{i}(x_{i}-\bar{x})^{\prime})$, where $\bar{x}=n^{-1}\sum_{i=1}^{n}x_i$. To facilitate our discussion on bias correction below, we present an alternative expression for $\hat{\tau}^{\mathrm{adj}}$. Let $z_{i}=(1,x_{i}^{\prime})^{\prime}$, where $\bar{x}$ is normalized to be zero for each coordinate, and $\hat{\beta}_{1}$ and $\hat{\beta}_{0}$ be the OLS estimators for the regression of $Y_{i}$ on $z_{i}$ by the treatment ($T_{i}=1$) and control ($T_{i}=0$) groups, respectively. Then the regression adjustment estimator can be written as
\[
\hat{\tau}^{\mathrm{adj}}=\hat{\mu}_{1}^{\mathrm{adj}}-\hat{\mu}_{0}^{\mathrm{adj}},\quad \mbox{where }\hat{\mu}_{t}^{\mathrm{adj}}=\frac{1}{n}\sum_{i=1}^{n}z_{i}^{\prime}\hat{\beta}_{t}\mbox{ for }t=0,1.
\]
 \citet*{lin2013agnostic} showed that $\hat{\tau}^{\mathrm{adj}}$ is consistent, asymptotically normal, and more efficient than the difference in means $\hat{\tau}^{\mathrm{dif}}$. It should be noted that these results hold true under the finite population setup with fixed $p$ without assuming correct specification of the linear model.

In practice, it is often the case that researchers observe many covariates. \citet*{lei2021regression} studied asymptotic properties of the regression adjustment estimator when the number of covariates $p$ grows with the sample size, and developed a bias-corrected estimator. To define Lei and Ding's (2021) approach, we introduce some notation. Let $Z=(z_{1},\ldots,z_{n})^{\prime}$, $P_{ij}$ be the $(i,j)$-th element of the projection matrix $P=Z(Z^{\prime}Z)^{-1}Z^{\prime}$, and $\hat{e}_{i}$ be the OLS residual, that is
\[
\hat{e}_{i} =
\begin{cases}
Y_{i}-z_{i}^{\prime}\hat{\beta}_{1} & \mbox{for treated units}\\
Y_{i}-z_{i}^{\prime}\hat{\beta}_{0} &  \mbox{for control units}
\end{cases}.
\]
Lei and Ding's (2021) bias-corrected estimator for $\tau$ is defined as 
\begin{equation}
\hat{\tau}^{\mathrm{bc}}=\left(\hat{\mu}_{1}^{\mathrm{adj}}+\frac{n_{0}}{n_{1}}\hat{\Delta}_{1}\right)-\left(\hat{\mu}_{0}^{\mathrm{adj}}+\frac{n_{1}}{n_{0}}\hat{\Delta}_{0}\right),\label{eq:bc}
\end{equation}
where
\[
\hat{\Delta}_{1}=n_{1}^{-1}\sum_{i=1}^{n}T_{i}P_{ii}\hat{e}_{i}, \quad \hat{\Delta}_{0}=n_{0}^{-1}\sum_{i=1}^{n}(1-T_{i})P_{ii}\hat{e}_{i}.
\]
Note that $\frac{n_{0}}{n_{1}}\hat{\Delta}_{1}$ and $\frac{n_{1}}{n_{0}}\hat{\Delta}_{0}$ are correction terms to estimate the higher-order bias terms of $\hat{\mu}_{1}^{\mathrm{adj}}$ and $\hat{\mu}_{0}^{\mathrm{adj}}$ under the moderate-$p$ asymptotics, respectively. The terms involving $\hat{\Delta}_{t}$ are analytic bias estimates that replace the unknown bias terms in their asymptotic theory. Although this bias correction method works in theory, the quality of these bias estimates may not be ideal, as illustrated in Section 4.4 of \citet*{lei2021regression}.

{  This paper studies an alternative bias correction approach via cross-fitting adapted from \citet{aronow-middleton2013}. We first note that the regression adjustment estimators for $\mu_{1}$ and $\mu_{0}$ can be alternatively written as 
\begin{align}
\hat{\mu}_{1}^{\mathrm{adj}}=\frac{1}{n}\sum_{i=1}^{n}\left\{ \frac{T_{i}}{\pi}Y_{i}-\left(\frac{T_{i}}{\pi}-1\right)(z_{i}^{\prime}\hat{\beta}_{1})\right\} ,\quad\hat{\mu}_{0}^{\mathrm{adj}}=\frac{1}{n}\sum_{i=1}^{n}\left\{ \frac{1-T_{i}}{1-\pi}Y_{i}-\left(\frac{1-T_{i}}{1-\pi}-1\right)(z_{i}^{\prime}\hat{\beta}_{0})\right\} ,\label{eq:ra}
\end{align}
where $\pi=n_{1}/n$ is the fraction of treated units treated as non-random in our setup.} Albeit the implementation differences, the estimation based on (\ref{eq:ra}) is equivalent to the full-sample regression adjustment estimation with treatment-covariate interactions first proposed by \citet*{lin2013agnostic} and the regression adjustment estimator in \citet*{lei2021regression}. The key idea of the bias correction is to replace the OLS estimators $\hat{\beta}_{1}$ and $\hat{\beta}_{0}$ with their leave-one-out counterparts
\begin{eqnarray*}
\hat{\beta}_{1}^{(i)} & = & \left(\sum_{j\neq i}T_{j}z_{j}z_{j}^{\prime}\right)^{-1}\left(\sum_{j\neq i}T_{j}z_{j}Y_{j}\right) \mbox{ for } i\in\{1,...,n:T_{i}=1\}, \\
\hat{\beta}_{0}^{(i)} & = & \left(\sum_{j\neq i}(1-T_{j})z_{j}z_{j}^{\prime}\right)^{-1}\left(\sum_{j\neq i}(1-T_{j})z_{j}Y_{j}\right) \mbox{ for } i\in\{1,...,n:T_{i}=0\}.
\end{eqnarray*}
Then the cross-fitted estimator of the average treatment effect $\tau$ is defined as
\[
\hat{\tau}^{\mathrm{cf}}=\hat{\mu}_{1}^{\mathrm{cf}}-\hat{\mu}_{0}^{\mathrm{cf}},
\]
where 
\begin{align}
\hat{\mu}_{1}^{\mathrm{cf}}=\frac{1}{n}\sum_{i=1}^{n}\left\{ \frac{T_{i}}{\pi}Y_{i}-\left(\frac{T_{i}}{\pi}-1\right)(z_{i}^{\prime}\hat{\beta}_{1}^{(i)})\right\} ,\quad\hat{\mu}_{0}^{\mathrm{cf}}=\frac{1}{n}\sum_{i=1}^{n}\left\{ \frac{1-T_{i}}{1-\pi}Y_{i}-\left(\frac{1-T_{i}}{1-\pi}-1\right)(z_{i}^{\prime}\hat{\beta}_{0}^{(i)})\right\} .\label{eq:cf}
\end{align}
Although this estimator may appear to be computationally demanding, in practice, one may utilize the identity for leave-one-out OLS estimation (see, e.g., Theorem 3.7 in \citet*{hansen2022econometrics}): 
\begin{equation}
\hat{\beta}_{t}^{(i)}=\hat{\beta}_{t}-(Z_{t}'Z_{t})^{-1}z_{i}\tilde{e}_{i}, \label{eq:betai}
\end{equation}
for $i\in\{1,...,n:T_{i}=t\}$, where $\tilde{e}_{i}=\hat{e}_{i}/(1-P_{t,ii})$ for $i$ with $T_i=t$, $P_{t,ij}$ is the $(i,j)$-th entry of the matrix $P_{t}=Z_{t}(Z_{t}'Z_{t})^{-1}Z_{t}'$, and $Z_{t}$ is the $n_{t}\times p$ submatrix that consists of $n_{t}$-rows of matrix $Z$ with $T_{j}=t$. This identity significantly lessens the computational burden to implement the cross-fitted estimator.

\section{Asymptotic theory}

In this section, we study asymptotic properties of the cross-fitted estimator $\hat{\tau}^{\mathrm{cf}}$ to compare with the existing ones, $\hat{\tau}^{\mathrm{adj}}$ and $\hat{\tau}^{\mathrm{bc}}$, and associated variance estimators.

\subsection{Stochastic expansion} \label{sec:main_bias_correction}

We first establish stochastic expansions for the estimators of $\tau$. To this end, we consider the setup employed by Lei and Ding (2021), where the number of covariates $p$ is allowed to grow with the sample size $n$. Let
\[
e_{i}(t)=y_{i}(t)-z_{i}^{\prime}\beta_{t},\quad \mbox{where }\beta_{t}=(Z^{\prime}Z)^{-1}Z^{\prime}Y(t) \mbox{ for }t=0,1,
\]
and
\[
\kappa=\max_{1\le i\le n}P_{ii},\quad \mathcal{E}_{2}=\max_{t\in\{0,1\}}\frac{1}{n}\sum_{i=1}^{n}e_{i}(t)^{2},\quad \mathcal{E}_{\infty}=\max_{t\in\{0,1\}}\max_{1\le i\le n}|e_{i}(t)|.
\]
We impose the following assumptions.

\begin{asm}\label{asm:1} $\quad$ 
\begin{description}
\item [{(i)}] {  $n_{1}$ and $n_{0}$ are non-random and satisfy $n/n_{1}=O(1)$ and $n/n_{0}=O(1)$.}
\item [{(ii)}] $\kappa\log p=o(1)$. 
\item [{(iii)}] $\sum_{i=1}^{n}e_{i}(1)e_{i}(0)/\sqrt{\sum_{i=1}^{n}e_{i}(1)^{2}\sum_{i=1}^{n}e_{i}(0)^{2}}>-1+\eta$,
for some constant $\eta>0$ independent of $n$. 
\item [{(iv)}] $\mathcal{E}_{\infty}^{2}/(n\mathcal{E}_{2})=o(1)$. 
\end{description}
\end{asm} Assumptions (i)-(iv) are identical to Assumptions 1-4 in \citet*{lei2021regression}, respectively. {  It should be noted that we impose no assumption on the functional forms of the outcome regression functions.} Assumption (i) holds if the proportions of treatment and control groups are fixed. 

{  Assumption (ii) restricts the growth rate of $p$. Since $\kappa \in [p/n,1]$, this assumption implies $p = o(n)$, i.e., $p$ should grow slower than $n$. Also this assumptions allows influential observations as long as their leverages are of smaller orders than $1/\log p$.} {  Note that $0 \le P_{ii} \le 1$ and $\sum^{n}_{i=1}P_{ii}=p$. Thus, in the favorable case where all leverage values are close to their average $p/n$, this condition holds if $\frac{p \log p}{n}=o(1)$.} Assumption (iii) imposes a mild restriction on the correlation between the potential residuals from the population ordinary least squares. It rules out perfectly negative correlation between the treatment and control potential residuals. Finally, Assumption (iv) imposes a Lindeberg-Feller type condition that none of potential residual dominates the others, while permitting heavy-tailed outcomes with $\mathcal{E}_{2}$ growing with $n$.

Let
\begin{eqnarray}
L_{i} & = & \left(\frac{T_{i}}{\pi}-1\right)e_{i}(1) - \left(\frac{1-T_{i}}{1-\pi}-1\right)e_{i}(0),\nonumber \\
W_{ij} & = & -\left(\frac{T_{i}}{\pi}-1\right)\left(\frac{T_{j}}{\pi}-1\right)z_{i}^{\prime}\Sigma^{-1}z_{j}\{e_{i}(1)+e_{j}(1)\}\nonumber \\
 &  & +\left(\frac{1-T_{i}}{1-\pi}-1\right)\left(\frac{1-T_{j}}{1-\pi}-1\right)z_{i}^{\prime}\Sigma^{-1}z_{j}\{e_{i}(0)+e_{j}(0)\}\nonumber \\
\mathcal{L} & = & \frac{1}{n}\sum_{i=1}^{n}L_{i}, \quad \mathcal{W} = \frac{1}{n^{2}}\sum_{1\le i<j\le n}W_{ij}, \quad \Sigma = \frac{1}{n}\sum^{n}_{i=1}z_i z_i^{\prime}. \label{eq:LW}
\end{eqnarray}
Under the above assumptions, higher-order asymptotic properties of the estimators for the ATE $\tau$ are obtained as follows.
{ 
\begin{thm}
\label{thm:bias} Consider the setup in Section \ref{sec:setup}, and suppose Assumptions (i)-(iv) hold true. 
\begin{description}
\item [{(i)}] Stochastic expansions of the estimators are obtained as
\begin{eqnarray}
\hat{\tau}^{\mathrm{adj}} - \tau  & = & B^{\mathrm{adj}} + \mathcal{L} + \mathcal{W} + O_{p}(n^{-1/2}\{\mathcal{E}_{2}(\kappa^2 p+\kappa^{3}p(\log p)^{2})\}^{1/2}), \nonumber \\
\hat{\tau}^{\mathrm{bc}} - \tau & = & \mathcal{L} + \mathcal{W} + O_{p}(n^{-1/2}\{\mathcal{E}_{2}(\kappa^{2}p+\kappa^{3}p(\log p)^{2})\}^{1/2}), \nonumber \\
\hat{\tau}^{\mathrm{cf}} - \tau & = & \mathcal{L} + \mathcal{W} + O_{p}(n^{-1/2}\{\mathcal{E}_{2}(\kappa^{2}p^{1/2}+\kappa^{3}p(\log p)^{2})\}^{1/2}), \label{eq:sto}
\end{eqnarray}
where the bias term of $\hat{\tau}^{\mathrm{adj}}$ is characterized as
\[
B^{\mathrm{adj}} = -\frac{1-\pi}{\pi}\frac{1}{n}\sum_{i=1}^{n}P_{ii}e_{i}(1)+\frac{\pi}{1-\pi}\frac{1}{n}\sum_{i=1}^{n}P_{ii}e_{i}(0)=O(n^{-1/2}({\mathcal E}_2\kappa p)^{1/2}).
\]
\item [{(ii)}] The first- and second-order stochastic terms satisfy 
\[
\mathcal{L}=O_{p}(n^{-1/2}\mathcal{E}_{2}^{1/2}),\qquad\mathcal{W}=O_{p}(n^{-1/2}(\mathcal{E}_{2}\kappa)^{1/2}),
\]
and their variances are characterized as
\begin{eqnarray*}
\sigma_{L}^{2} & := & \mathbb{V}(\sqrt{n}\mathcal{L}) \\
 & = & \frac{n}{n_{1}(n-1)}\sum_{i=1}^{n}e_{i}(1)^{2}+\frac{n}{n_{0}(n-1)}\sum_{i=1}^{n}e_{i}(0)^{2}-\frac{1}{n-1}\sum_{i=1}^{n}(e_{i}(1)-e_{i}(0))^{2}\\
 & = & O(\mathcal{E}_{2}),
\end{eqnarray*}
\begin{eqnarray*}
\sigma_{W}^{2} & := & \mathbb{V}(\sqrt{n}\mathcal{W}) \\
 & = & \frac{n_{0}^{2}}{n_{1}^{2}n}\sum_{i=1}^{n}P_{ii}e_{i}(1)^{2}+\frac{n_{1}^{2}}{n_{0}^{2}n}\sum_{i=1}^{n}P_{ii}e_{i}(0)^{2}-\frac{2}{n}\sum_{i=1}^{n}P_{ii}e_{i}(1)e_{i}(0)\\
 &  & +\frac{n_{0}^{2}}{n_{1}^{2}n}\sum_{i=1}^{n}\sum_{j\neq i}^{n}P_{ij}^{2}\left\{ -e_{i}(1)+\left(\frac{\pi}{1-\pi}\right)^{2}e_{i}(0)\right\} \left\{ -e_{j}(1)+\left(\frac{\pi}{1-\pi}\right)^{2}e_{j}(0)\right\} \\
 & = & O(\mathcal{E}_{2}\kappa).
\end{eqnarray*}
\end{description}
\end{thm}
}

Theorem \ref{thm:bias} (i) decomposes the estimation errors into a first-order dominant linear term $\mathcal{L}$, a second-order quadratic term $\mathcal{W}$, and the bias terms $B^{\mathrm{adj}}$ for $\hat{\tau}^{\mathrm{adj}}$. Note that the linear and quadratic terms are identical for all the estimators, and the differences are attributed to the bias term and stochastic orders of the remainder terms.

The bias term $B^{\mathrm{adj}}$ for the conventional regression adjustment estimator is studied by \citet*{lei2021regression}. In contrast, Lei and Ding's (2021) bias-corrected estimator $\hat{\tau}^{\mathrm{bc}}$ and the cross-fitted estimator $\hat{\tau}^{\mathrm{cf}}$ do not involve such a bias term and have better higher-order bias properties. Since the stochastic term $\mathcal{L}+\mathcal{W}$ is identical for all estimators, such bias reducing features of $\hat{\tau}^{\mathrm{bc}}$ and $\hat{\tau}^{\mathrm{cf}}$ do not inflate the variance compared to the conventional regression adjustment estimator. {  However, our theorem does not necessarily mean $\hat{\tau}^{\mathrm{cf}}$ is exactly unbiased: the remainder term in (\ref{eq:sto}) is typically small but non-zero.} Furthermore, compared to \citet*{lei2021regression}, our expansions also characterize the second order quadratic term $\mathcal{W}$, which will be useful to investigate higher-order properties of the variance estimators in the next subsection.\footnote{  One may also consider a leave-one-out version of the regression adjustment estimator $\tilde{\tau}^{\mathrm{adj}} = \tilde{\mu}_{1} - \tilde{\mu}_{0}$, where $\tilde{\mu}_{t} = n^{-1}\sum_{i=1}^{n}z^{\prime}_{i}\hat{\beta}^{(i)}_{t}$. However, this estimator possesses an analogous bias term to the regression adjustment estimator. To see this, the identity in (\ref{eq:betai}) and an analogous argument in the proof of Theorem \ref{thm:bias} yield
\[
\tilde{\mu}_{1} = \hat{\mu}^{\mathrm{cf}}_{1} - \frac{1}{n}\sum_{i=1}^n \frac{T_i}{\pi}z_i^{\prime}(Z_1^{\prime}Z_1)^{-1}z_i\tilde{e}_i = \left\{\hat{\mu}^{\mathrm{cf}}_{1} - \frac{1}{n\pi}\sum_{i=1}^n \frac{T_i}{\pi}P_{ii}e_i(1)\right\}(1+o_p(1)).
\]
Thus, the bias term of $\tilde{\tau}^{\mathrm{adj}}$ will be of same order as $B^{\mathrm{adj}}$.}



We now compare $\hat{\tau}^{\mathrm{bc}}$ and $\hat{\tau}^{\mathrm{cf}}$. As shown in Theorem \ref{thm:bias} (i), the remainder terms of these exhibit different orders. Also Theorem \ref{thm:bias} (ii) characterizes the stochastic components $\mathcal{L}$ and $\mathcal{W}$. Combining these results, the stochastic orders of the estimation errors are
\begin{align*}
\hat{\tau}^{\mathrm{adj}}-\tau &=O_{p}(n^{-1/2}\{\mathcal{E}_{2}(\kappa p+1)\}^{1/2}),\\
\hat{\tau}^{\mathrm{bc}}-\tau &=O_{p}(n^{-1/2}\{\mathcal{E}_{2}(\kappa^2 p+1)\}^{1/2}),\\
\hat{\tau}^{\mathrm{cf}}-\tau &=O_{p}(n^{-1/2}\{\mathcal{E}_{2}(\kappa^2p^{1/2}+\kappa^3 p (\log p)^2+1)\}^{1/2}).
\end{align*}
{ We note that $\mathcal{L}$ is asymptotically normal with mean $0$ and variance $\sigma_L^2/n$; see Lei and Ding (2021, Theorem 3). Hence, for $a\in\{\mathrm{adj},\mathrm{bc},\mathrm{cf}\}$, the estimator $\hat{\tau}^{a}$ has the same asymptotic normality as $\mathcal{L}$ if the remainder terms (as well as the bias term $B^{\mathrm{adj}}$ for $\hat{\tau}^{\mathrm{adj}}$) vanish when multiplied by $\sqrt{n}/\sigma_L$. Specifically, the convergence $\sqrt{n}(\hat{\tau}^{a}-\tau)/\sigma_{L}\overset{d}{\to}N(0,1)$ holds if $\kappa p=o(1)$ for $a=\mathrm{adj}$, if $\kappa^2 p=o(1)$ for $a=\mathrm{bc}$, and if $\kappa^2 p^{1/2}+\kappa^3p(\log p)^2=o(1)$ for $a=\mathrm{cf}$, respectively. In the favorable case where all leverage scores are close to their average $p/n$, the regression adjustment estimator and Lei and Ding's bias corrected estimator are asymptotically normal when $p=o(n^{1/2})$ and $p=o(n^{2/3})$, respectively, but the cross-fitted estimator is asymptotically normal when $p=o(n^{3/4}/(\log n)^{1/2})$.}

In addition to the linear component $\mathcal{L}$, Theorem \ref{thm:bias} (ii) characterizes the variance of the second-order quadratic term $\mathcal{W}$. The term $\sigma_{L}^{2}$ is identical to the conventional variance term for the regression adjustment estimator as in Lin (2013). {  Therefore, similar to the regression adjustment and Lei and Ding's (2021) bias-corrected estimators, the cross-fitted estimator is guaranteed to be asymptotically more efficient than the difference in means estimator despite arbitrary misspecification of the conditional mean function.} Note that the third component in the expression of $\sigma_{L}^{2}$, $(n-1)^{-1}\sum_{i=1}^{n}(e_{i}(1)-e_{i}(0))^{2}$, has no consistent estimator in general. The additional term $\sigma_{W}^{2}$ also contains a component which cannot be consistently estimated (i.e., the third term of $\sigma_{W}^{2}$).

Compared to the existing results such as \citet*{lei2021regression}, the results on the second-order term $\mathcal{W}$ and its variance $\sigma_{W}^{2}$ are new. Indeed, in their simulation study, Lei and Ding (2021) reported that $\sigma_{L}^{2}$ tends to be lower than the Monte Carlo variance of the point estimator for $\tau$ for larger values of $p$. Based on our higher-order analysis, we argue that this discrepancy can be attributed to the second-order component $\sigma_{W}^{2}$ whose order increases with $p$.

\subsection{Variance estimation}

We next consider variance estimation of the treatment effect estimator, particularly the HC0 and HC3 variance estimators 
\begin{eqnarray*}
\hat{\sigma}_{\text{HC0}}^{2} & = & \frac{n}{n_{1}(n_{1}-1)}\sum_{i=1}^{n}T_{i}\hat{e}_{i}^{2}+\frac{n}{n_{0}(n_{0}-1)}\sum_{i=1}^{n}(1-T_{i})\hat{e}_{i}^{2},\\
\hat{\sigma}_{\text{HC3}}^{2} & = & \frac{n}{n_{1}(n_{1}-1)}\sum_{i=1}^{n}T_{i}\tilde{e}_{i}^{2}+\frac{n}{n_{0}(n_{0}-1)}\sum_{i=1}^{n}(1-T_{i})\tilde{e}_{i}^{2}.
\end{eqnarray*}
{  Lei and Ding (2021, Theorem 5) showed that under Assumptions (i)-(iv), both estimators are asymptotically conservative, i.e., $\hat{\sigma}_{\text{HC}j}^{2}/\sigma^{2}_{L}\ge 1-a_{jn}$ with a non-negative sequence $a_{jn}=o_p(1)$ for $j=0$ and $3$.} Under our setup, the properties of these variance estimators are characterized as follows.
{ 
\begin{thm}
\label{thm:var1} Consider the setup in Section \ref{sec:setup}. Suppose Assumptions (i)-(iv) hold true. The variance estimators satisfy 
\begin{align*}
\hat{\sigma}_{\rm{HC0}}^{2} &= \left\{\frac{n}{n_{1}(n-1)}\sum_{i=1}^{n}e_{i}(1)^{2}+\frac{n}{n_{0}(n-1)}\sum_{i=1}^{n}e_{i}(0)^{2}-\frac{n_{0}}{n_{1}^{2}}\sum_{i=1}^{n}P_{ii}e_{i}(1)^{2}-\frac{n_{1}}{n_{0}^{2}}\sum_{i=1}^{n}P_{ii}e_{i}(0)^{2}\right\}(1+o_p(1)),\\
\hat{\sigma}_{\rm{HC3}}^{2} &= \left\{\frac{n}{n_{1}(n-1)}\sum_{i=1}^{n}e_{i}(1)^{2}+\frac{n}{n_{0}(n-1)}\sum_{i=1}^{n}e_{i}(0)^{2}+\frac{n_{0}}{n_{1}^{2}}\sum_{i=1}^{n}P_{ii}e_{i}(1)^{2}+\frac{n_{1}}{n_{0}^{2}}\sum_{i=1}^{n}P_{ii}e_{i}(0)^{2}\right\}(1+o_p(1)).
\end{align*}
\end{thm}
}

This theorem depicts the means of the HC0 and HC3 variance estimators, taking into account of the higher-order terms.
First, the first two terms of $\mathbb{E}[\hat{\sigma}_{\text{HC0}}^{2}]$ and $\mathbb{E}[\hat{\sigma}_{\text{HC3}}^{2}]$ are the exact match to the first two terms of $\sigma_{L}^{2}$. However, the third term of $\sigma_{L}^{2}$ is not consistently estimable. Thus, as far as we are concerned with the first-order dominant terms, HC0 and HC3 are conservative estimators of the asymptotic variance of the treatment effect estimators. Second, the third and fourth terms of $\mathbb{E}[\hat{\sigma}_{\text{HC3}}^{2}]$ closely match to the first and second terms of $\sigma_{W}^{2}$, except for the factors $n_{0}/n$ and $n_{1}/n$, respectively. It is interesting to note that the HC3 estimator is interpreted as a jackknife variance estimator. So these multiplicative discrepancies can be understood as emergence of Efron and Stein's (1981) bias for the jackknife variance in higher-order terms in the context of the design-based asymptotic analysis. Third, it should be noted that the signs of the third and fourth terms of $\mathbb{E}[\hat{\sigma}_{\text{HC0}}^{2}]$ are opposite to the corresponding ones in the first and second terms of $\sigma_{W}^{2}$ (or the signs of the third and fourth terms of $\mathbb{E}[\hat{\sigma}_{\text{HC3}}^{2}]$). Therefore, the higher-order term of HC3 slightly overestimates $\sigma_{W}^{2}$, while HC0 severely underestimates $\sigma_{W}^{2}$. This explains relatively poor performances of HC0 in finite samples, as observed in the literature (e.g., simulation studies in \citet*{lei2021regression}).

{  Note that all the terms of $\sigma_W^2$ except the third term are estimable. It is of interest whether one can construct a variance estimator that is guaranteed to be asymptotically conservative for the variance component $\sigma^2_L + \sigma^2_W$ up to the second-order. To this end, we propose a modified version of the HC3 variance estimator:
\begin{eqnarray*}
\hat{\sigma}_{\text{mHC3 }}^{2} & = &  \hat{\sigma}_{\text{HC3}}^{2} + \frac{n_1-n_0}{n_1^2}\sum_{i=1}^n T_iP_{ii}\tilde{e}_i^2 + \frac{n_0-n_1}{n_0^2}\sum_{i=1}^n (1-T_i)P_{ii}\tilde{e}_i^2\\
& & +\frac{n_{0}^{2}n}{n_{1}^{4}}\sum_{i=1}^{n}\sum_{j\neq i}^{n}P_{ij}^{2}T_{i}T_{j}\tilde{e}_{i}\tilde{e}_{j}+\frac{n_{1}^{2}n}{n_{0}^{4}}\sum_{i=1}^{n}\sum_{j\neq i}^{n}P_{ij}^{2}(1-T_{i})(1-T_{j})\tilde{e}_{i}\tilde{e}_{j} \\
 & & -\frac{2n}{n_{0}n_{1}}\sum_{i=1}^{n}\sum_{j\neq i}^{n}P_{ij}^{2}T_{i}(1-T_{j})\tilde{e}_{i}\tilde{e}_{j}.
\end{eqnarray*}
The asymptotic conservativeness of $\hat{\sigma}_{\text{mHC3 }}^{2}$ is obtained as follows.
\begin{cor}\label{cor:mHC3}
Consider the setup in Section \ref{sec:setup}. Suppose Assumptions (i)-(iv) hold true. Then there exists a non-negative sequence $a_n=o_p(1)$ such that $\hat{\sigma}_{\rm{mHC3 }}^{2}/(\sigma^2_L + \sigma^2_W) \ge 1-a_n$.
\end{cor}
}
We investigate its finite sample performance in the simulation study below.\footnote{  By estimating the estimable components in $\sigma_W^2$, we can propose alternative bias-corrected versions of the HC0 and HC3 variance estimators defined as follows:
\begin{eqnarray*}
\hat{\sigma}_{\text{dbHC0}}^{2} & = &  \hat{\sigma}_{\text{HC0}}^{2} +\frac{n_0(n_0+n)}{n_1^3}\sum_{i=1}^n T_iP_{ii}\tilde{e}_i^2 +\frac{n_1(n_1+n)}{n_0^3}\sum_{i=1}^n (1-T_i)P_{ii}\tilde{e}_i^2\\
& & +\frac{n_{0}^{2}n}{n_{1}^{4}}\sum_{i=1}^{n}\sum_{j\neq i}^{n}P_{ij}^{2}T_{i}T_{j}\tilde{e}_{i}\tilde{e}_{j}+\frac{n_{1}^{2}n}{n_{0}^{4}}\sum_{i=1}^{n}\sum_{j\neq i}^{n}P_{ij}^{2}(1-T_{i})(1-T_{j})\tilde{e}_{i}\tilde{e}_{j} -\frac{2n}{n_{0}n_{1}}\sum_{i=1}^{n}\sum_{j\neq i}^{n}P_{ij}^{2}T_{i}(1-T_{j})\tilde{e}_{i}\tilde{e}_{j}. \\
\hat{\sigma}_{\text{dbHC3 }}^{2} & = &  \hat{\sigma}_{\text{HC3}}^{2} -\frac{n_0}{n_1^2}\sum_{i=1}^n T_iP_{ii}\tilde{e}_i^2 -\frac{n_1}{n_0^2}\sum_{i=1}^n (1-T_i)P_{ii}\tilde{e}_i^2\\
& & +\frac{n_{0}^{2}n}{n_{1}^{4}}\sum_{i=1}^{n}\sum_{j\neq i}^{n}P_{ij}^{2}T_{i}T_{j}\tilde{e}_{i}\tilde{e}_{j}+\frac{n_{1}^{2}n}{n_{0}^{4}}\sum_{i=1}^{n}\sum_{j\neq i}^{n}P_{ij}^{2}(1-T_{i})(1-T_{j})\tilde{e}_{i}\tilde{e}_{j} -\frac{2n}{n_{0}n_{1}}\sum_{i=1}^{n}\sum_{j\neq i}^{n}P_{ij}^{2}T_{i}(1-T_{j})\tilde{e}_{i}\tilde{e}_{j}.
\end{eqnarray*}
However, analogous arguments to the proof of Corollary \ref{cor:mHC3} show that these asymptotic variance estimators do not achieve asymptotic conservativeness up to the second-order.}



\section{Stratified randomized experiments with large strata}
Stratified randomized experiments using regression adjustment estimators have been considered in \cite{liu2020regression} under a fixed dimensional asymptotic regime. Our methodology can be extended to the stratified randomized experiments with a finite number of large strata. Consider stratified randomized experiments with $N$ units in the population grouped into strata $s=1,\ldots,S$ for a finite $S$. For each stratum, a randomized experiment is then conducted independently from other strata. The size of the $s$-th stratum is denoted as $n_s\ge 2$. Within the stratum, $n_{1s}$ of them are sampled without replacement and receive treatment while the remaining $n_{0s}=n_{s}-n_{1s}$ are assigned to control. Let $\pi_s=n_{1s}/(n_{1s}+n_{0s})$. Denote the potential outcomes of unit $i$ in stratum $s$ as $(y_{is}(1),y_{is}(0))$, and $T_{is}$, $Y_{is}$, and $z_{is}$ are the corresponding observed outcome, treatment indicator variable, and vector of covariates, respectively. The population average treatment effect is defined as
\[
\tau=\sum_{s=1}^S\sum_{i=1}^{n_s} \frac{\tau_{is}}{N}=\sum_{s=1}^S c_s\tau_s,
\]
where $\tau_{is}=y_{is}(1)-y_{is}(0)$, $\tau_s=\sum_{i=1}^{n_s}\tau_{is}/n_s$, and $c_s=n_s/N$. For each unit $i$ in stratum $s$, define the leave-one-out estimators $\hat \beta_{1s}^{(j)}=\left(\sum_{i\in I_s \setminus \{j\}}T_{is}z_{is}z_{is}^{\prime}\right)^{-1}\left(\sum_{i\in I_s \setminus \{j\}}T_{is}z_{is}Y_{is}\right)$, and $\hat \beta_{0s}^{(j)}=\left(\sum_{i\in I_s \setminus \{j\}}(1-T_{is})z_{is}z_{is}^{\prime}\right)^{-1}\left(\sum_{i\in I_s \setminus \{j\}}(1-T_{is})z_{is}Y_{is}\right)$, where $I_s$ means the units in stratum $s$. The cross-fitted estimator for $\tau$ is then defined as $\hat\tau = \sum_{s=1}^S \sum_{i=1}^{n_s} \hat\tau_s/N$, where $\hat\tau_s=\hat \mu_{1s}^{\text{cf}}-\hat \mu_{0s}^{\text{cf}}$,
\[
\hat{\mu}_{1s}^{\mathrm{cf}}=\frac{1}{n_s}\sum_{j\in I_s}\left\{ \frac{T_{js}}{\pi_s}Y_{js}-\left(\frac{T_{js}}{\pi_s}-1\right)(z_{js}^{\prime}\hat{\beta}_{1s}^{(j)})\right\},\quad\hat{\mu}_{0s}^{\mathrm{cf}}=\frac{1}{n_s}\sum_{j\in I_s}\left\{ \frac{1-T_{js}}{1-\pi_s}Y_{js}-\left(\frac{1-T_{js}}{1-\pi_s}-1\right)(z_{js}^{\prime}\hat{\beta}_{0s}^{(j)})\right\}.
\]
Then, as long as Assumptions \ref{asm:1} (i)-(iv) hold for each stratum $s$, as $\min_{s=1,\ldots,S} n_s$ diverges to infinity, the conclusions of Theorems \ref{thm:bias}-\ref{thm:var1} continue to hold for each stratum. As the estimators $\hat \tau_1,\ldots,\hat \tau_S$ are mutually independent, the variance of $\hat\tau$ can be estimated by
{ 
\begin{align*}
\hat \sigma_{\text{HC3}}^2=&\sum_{s=1}^S c_s^2 \left\{\frac{n_s}{n_{1s}(n_{1s}-1)}\sum_{i\in I_s}T_{is} \tilde e_{is}^2+ \frac{n_s}{n_{0s}(n_{0s}-1)}\sum_{i\in I_s}(1-T_{is}) \tilde e_{is}^2\right\},\\
\hat \sigma_{\text{mHC3}}^2=&\hat\sigma_{\text{HC3}}^2 + \sum_{d=1}^S c_s^2
\Bigg\{\frac{n_{1s}-n_{0s}}{n_{1s}^2}\sum_{i\in I_s}T_{is}P_{ii,s}\tilde e_{is}^2+ \frac{n_{0s}-n_{1s}}{n_{0s}^2}\sum_{i\in I_s}(1-T_{is})P_{ii,s}\tilde e_{is}^2\\
&+\frac{n_{0s}^2 n_s}{n_{1s}^4}\sum_{i\in I_s}\sum_{j\in I_s\setminus\{i\}}P_{ij,s}^2T_{is}T_{js}\tilde e_{is}\tilde e_{js}+ \frac{n_{1s}^2n_s}{n_{0s}^4}\sum_{i\in I_s}\sum_{j\in I_s\setminus\{i\}}P_{ij,s}^2(1-T_{is})(1-T_{js})\tilde e_{is}\tilde e_{js} 
\\&- \frac{2n_s}{n_{0s}n_{1s}} \sum_{i\in I_s}\sum_{j\in I_s\setminus\{i\}} P_{ij,s}^2T_{is}(1-T_{js})\tilde e_{is}\tilde e_{js} \Bigg\},
\end{align*}
}
where $\tilde e_{is}$ and $P_{ij,s}$ are defined in the same way as $\tilde e_{i}$, $P_{ij}$, respectively, with observations restricted to the $s$-th stratum. Asymptotically valid inference under this stratified setup can be conducted based on these variance estimators.

\section{Simulation}

In this section, we illustrate our theoretical findings through a series of simulation studies. The simulation designs closely follow those in \citet*{lei2021regression}. Specifically, we set the sample size to \( n = 500 \) and define the treatment group size as \( n_{1} = n\pi_{1} \) with \( \pi_{1} = 0.2 \). The covariate matrix \( \mathcal{X} \in \mathbb{R}^{n \times n} \) consists of independent and identically distributed entries drawn from the $t(3)$ distribution. To mimic the design-based asymptotic framework, the matrix \( \mathcal{X} \) is generated once and held fixed across all Monte Carlo replications. Similarly, a vector \( b \in \mathbb{R}^n \), with entries iid from a standard normal distribution, is generated once at the beginning and subsequently kept fixed.

For each dimension \( p \in \{5, 10, \dots, 75\} \), we construct the covariate matrix \( X \in \mathbb{R}^{n \times p} \) by taking the first \( p \) columns of \( \mathcal{X} \), and define parameter vectors \( \beta_{1}^{*} = \beta_{0}^{*} = (b_1, \dots, b_p)^{\prime} \) using the first \( p \) entries of \( b \). The potential outcomes are specified as
\[
Y(t) = X \beta_t^* + \epsilon(t), \quad t \in \{0, 1\},
\]
where the error vectors \( \epsilon(t) \in \mathbb{R}^n \) are generated under two distinct designs: a worst-case configuration and a normal-error design.

For the worst-case errors, we define \( \epsilon(0) = \epsilon \) and \( \epsilon(1) = 2\epsilon \), where the vector \( \epsilon \in \mathbb{R}^n \) solves the constrained optimization problem:
\[
\max_{\epsilon \in \mathbb{R}^n} \left| \frac{n_1}{n_0} \Delta_0 - \frac{n_0}{n_1} \Delta_1 \right| \quad \text{subject to } \frac{1}{n} \epsilon^{\prime} \epsilon = 1, \quad X^{\prime} \epsilon = (1,...,1)^{\prime} \epsilon = 0,
\]
with \( \Delta_t = n^{-1} \sum_{i=1}^n e_i(t) P_{ii} \). This construction maximizes the first-order bias of the regression-adjusted estimator \( \hat{\tau}^{\mathrm{adj}} \) under the increasing-dimensional asymptotics developed by \citet*{lei2021regression}.
For the normal error design, we consider homoskedastic normal errors with \( \epsilon(0) = \epsilon(1) = \epsilon \), where \( \epsilon \sim N(0, I_n) \). In this setting, the potential outcome equations are linear, and biases from regression adjustment are generally small. As with the covariates \( \mathcal{X} \), the error vectors \( \epsilon(t) \) are generated once and fixed throughout the simulation replications.
Each simulation design is evaluated with $10,000$ Monte Carlo repetitions.

We compare three estimators for the average treatment effect of interest:
(i) the standard regression adjustment estimator based on equation~(\ref{eq:ra}) (RA),
(ii) the bias-corrected regression adjustment estimator from \citet*{lei2021regression}, given in equation~(\ref{eq:bc}) (BC), and
(iii) the cross-fitted regression adjustment estimator introduced in this paper, defined in equation~(\ref{eq:cf}) (CF).
For all estimators, inference is conducted using heteroskedasticity-robust standard errors from the Eicker-Huber-White family, specifically HC2 and HC3. As reported in \citet*{lei2021regression}, HC3 tends to yield the most reliable performance across simulation settings, particularly when the covariate dimension is high. For CF, we further consider inference based on the modified HC3 variance estimator (mHC3), proposed in this paper. The mHC3 estimator is designed to provide improved higher-order accuracy in the context of cross-fitting, potentially offering more robust inference than standard HC3.

\begin{figure}[!t]
	\centering
	
	\includegraphics[width=1\textwidth]{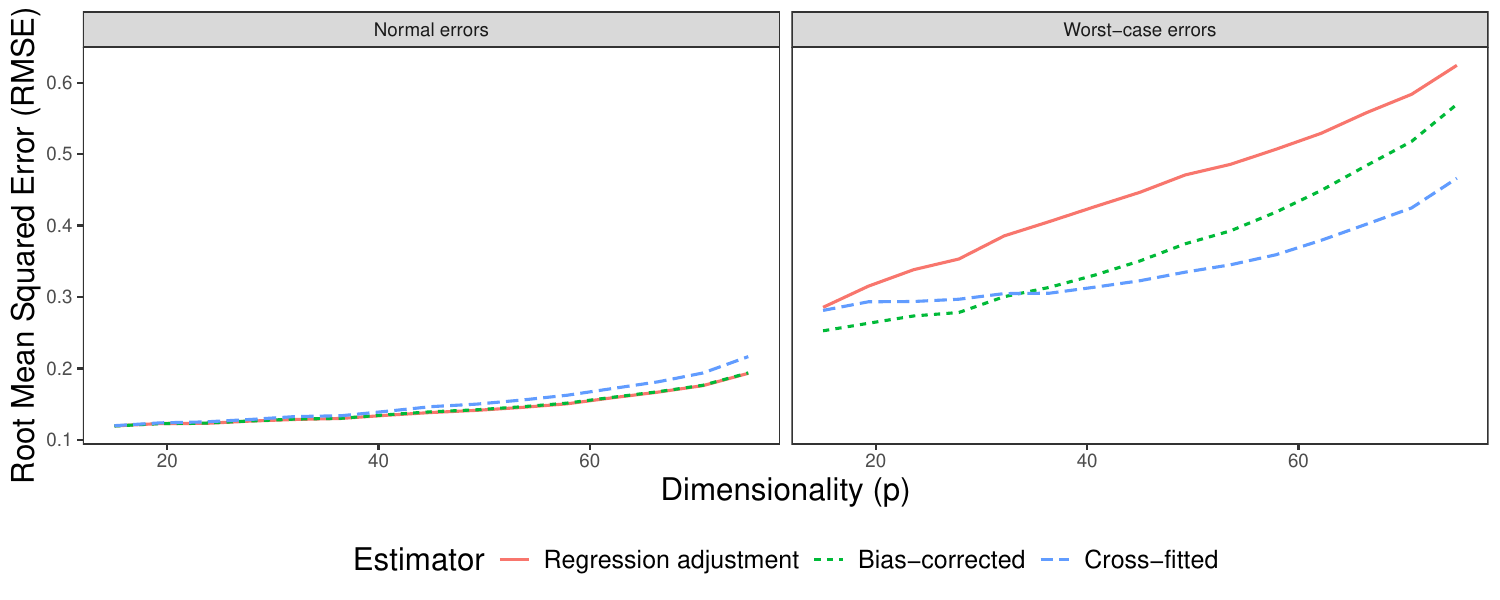}
	\caption{RMSE Comparison of estimators}
\label{fig:RMSE}
\end{figure}

Figure \ref{fig:RMSE} shows the root mean squared errors (RMSE) for the three estimators under normal and worst-case errors. Note that when the errors are normal, all three estimators performed decently. The differences become much more pronounced under the worst-case error structure. RA suffers severe degradation in performance as $p$ increases. Its RMSE rises sharply with $p$, underscoring its vulnerability to adversarial alignment between the error and the design. BC successfully controls this bias, maintaining relatively low RMSE throughout. However, its RMSE still shows mild growth in higher dimensions. CF delivers the best performance, and consistently outperforms RA and BC, and the performance gap widens with increasing  $p$, affirming the theoretical advantages of cross-fitting in high-dimensional and misspecified environments.

\begin{figure}[!t]
	\centering
	
	\includegraphics[width=1\textwidth]{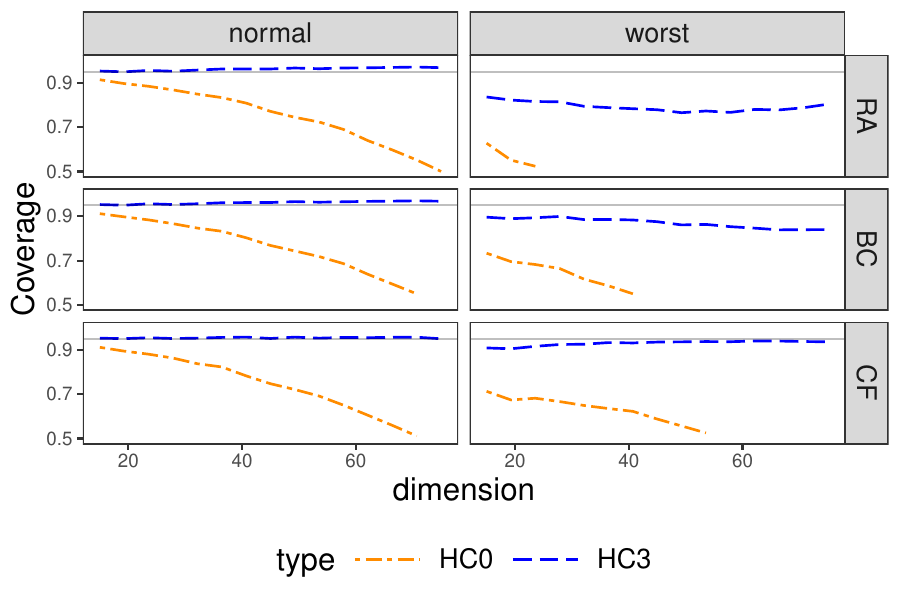}
	\caption{Coverage rates for the three estimators with HC2 and HC3}
\label{fig:coverage_1}
\end{figure}

\begin{figure}[!t]
	\centering
	
	\includegraphics[width=1\textwidth]{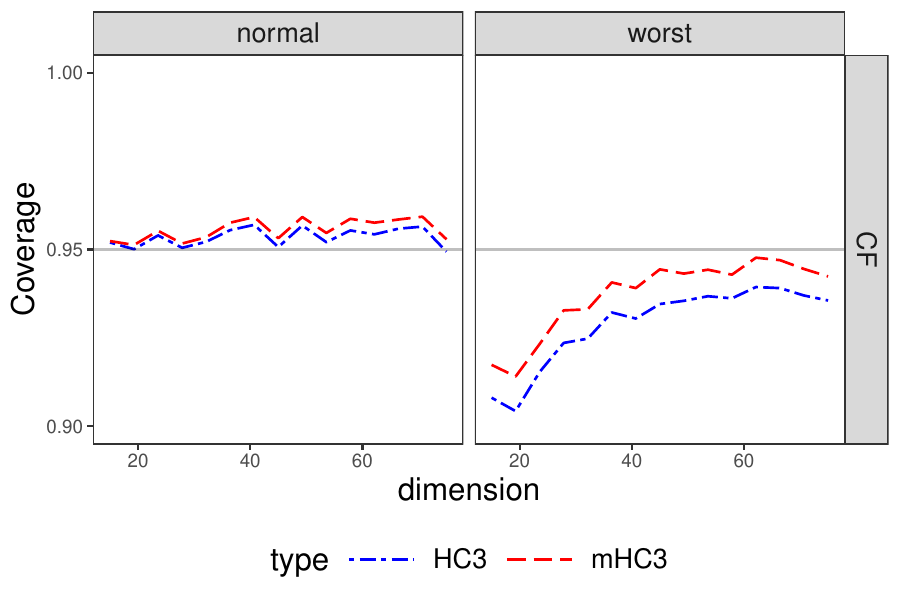}
	\caption{Coverage rates for cross-fitted estimator with HC2, HC3, and mHC3}
\label{fig:coverage_2}
\end{figure}

Figure \ref{fig:coverage_1} shows the coverage rates for the three estimators with different variance estimators. Observe that HC2 fails to maintain nominal coverage as dimension increases. Coverage drops steadily with $p$, indicating that HC2 underestimates the variability of the estimator in challenging settings. HC3 performs better than HC2 but still exhibits noticeable undercoverage for larger values of $p$. Figure \ref{fig:coverage_2} focuses on the CF estimator and highlights the contrast in coverage rates between HC2, HC3, and mHC3. The performance gap between mHC3 and the conventional estimators grows with $p$, showcasing the benefit of incorporating bias correction in the standard error estimator when dealing with highly misspecified or adversarial settings.

{  In sum, we recommend to use the cross-fitted estimator for point estimation of the average treatment effect, and the HC3 or mHC3 variance estimator for interval estimation to practitioners particularly when the number of covariates $p$ is moderate or large. Theoretically, the cross-fitted estimator is asymptotically more efficient than the unbiased difference in means estimator, and admits asymptotic normality under weaker conditions on $p$ compared to the regression adjustment and Lei and Ding's (2021) bias-corrected estimators.}

\newpage{}

\appendix

\section{Mathematical appendix}

{ 
\textbf{Notation:} Hereafter, let  $A_{n}\lesssim B_{n}$ means that there exists a positive constant $C$ independent of $n$ such that $A_{n}\le CB_{n}$ for all $n$ a.s., $A_{n}\gtrsim B_{n}$ means $B_{n}\lesssim A_{n}$, and $A_{n}\sim B_{n}$ means $A_{n}\lesssim B_{n}$ and $B_{n}\lesssim A_{n}$.\\ 
}

In this appendix we use the following notation. Let 
\begin{eqnarray*}
\Sigma & = & \frac{1}{n}\sum_{i=1}^{n}z_{i}z_{i}^{\prime},=\frac{1}{n}Z^{\prime}Z\qquad\Sigma_{1}=\frac{1}{n_{1}}\sum_{i=1}^{n}T_{i}z_{i}z_{i}^{\prime},\qquad\Sigma_{1}^{(i)}=\frac{1}{n_{1}}\sum_{j\neq i}^{n}T_{j}z_{j}z_{j}^{\prime},\\
\Sigma_{0} & = & \frac{1}{n_{0}}\sum_{i=1}^{n}(1-T_{i})z_{i}z_{i}^{\prime},\qquad\Sigma_{0}^{(i)}=\frac{1}{n_{0}}\sum_{j\neq i}^{n}(1-T_{j})z_{j}z_{j}^{\prime},\\
\rho_{i} & = & -e_{i}(1)+\left(\frac{\pi}{1-\pi}\right)^{2}e_{i}(0),\qquad r_{i}=e_{i}(1)+\left(\frac{\pi}{1-\pi}\right)^{3}e_{i}(0),
\end{eqnarray*}
We repeatedly use the following facts. Since $e_{i}(t)$ is the OLS
residual, it holds 
\begin{equation}
\sum_{i=1}^{n}z_{i}e_{i}(t)=0,\qquad\sum_{i=1}^{n}x_{i}e_{i}(t)=0,\qquad\sum_{i=1}^{n}e_{i}(t)=0,\label{pf:e}
\end{equation}
for $t=0,1$. Also the projection matrix $P=Z(Z^{\prime}Z)^{-1}Z^{\prime}$
satisfies 
\begin{equation}
\sum_{i=1}^{n}P_{ii}=p,\qquad\sum_{j=1}^{n}P_{ij}=1,\qquad\sum_{j=1}^{n}P_{ij}^{2}=P_{ii},\qquad\sum_{j=1}^{n}P_{ij}P_{jk}=P_{ik},\qquad P_{ii}<1.\label{pf:P}
\end{equation}
Finally, we note that 
\begin{equation}
\frac{1-T_{i}}{1-\pi}-1=-\frac{\pi}{1-\pi}\left(\frac{T_{i}}{\pi}-1\right).\label{pf:T}
\end{equation}

\subsection{Proof of Theorem \ref{thm:bias}}

\subsubsection{Proof of (i)}

First, we consider $\hat{\tau}^{\mathrm{adj}}=\hat{\mu}_{1}^{\mathrm{adj}}-\hat{\mu}_{0}^{\mathrm{adj}}$. By (\ref{pf:e}) and (\ref{pf:T}), we decompose 
\begin{eqnarray}
 &  & \hat{\tau}^{\mathrm{adj}}-\tau = (\hat{\mu}_{1}^{\mathrm{adj}}-\mu_{1})-(\hat{\mu}_{0}^{\mathrm{adj}}-\mu_{0})\nonumber \\
 & = & \frac{1}{n}\sum_{i=1}^{n}\left(\frac{T_{i}}{\pi}-1\right)\{e_{i}(1)-z_{i}^{\prime}(\hat{\beta}_{1}-\beta_{1})\}-\frac{1}{n}\sum_{i=1}^{n}\left(\frac{1-T_{i}}{1-\pi}-1\right)\{e_{i}(0)-z_{i}^{\prime}(\hat{\beta}_{0}-\beta_{0})\}\nonumber \\
 & = & \frac{1}{n}\sum_{i=1}^{n}\left(\frac{T_{i}}{\pi}-1\right)\left(e_{i}(1)+\frac{\pi}{1-\pi}e_{i}(0)\right)-\left\{ \frac{1}{n}\sum_{i=1}^{n}\left(\frac{T_{i}}{\pi}-1\right)z_{i}^{\prime}\right\} \Sigma_{1}^{-1}\left\{ \frac{1}{n}\sum_{i=1}^{n}\left(\frac{T_{i}}{\pi}-1\right)z_{i}e_{i}(1)\right\} \nonumber \\
 &  & +\left(\frac{\pi}{1-\pi}\right)^{2}\left\{ \frac{1}{n}\sum_{i=1}^{n}\left(\frac{T_{i}}{\pi}-1\right)z_{i}^{\prime}\right\} \Sigma_{0}^{-1}\left\{ \frac{1}{n}\sum_{i=1}^{n}\left(\frac{T_{i}}{\pi}-1\right)z_{i}e_{i}(0)\right\} \nonumber \\
 & =: & M_{1}+M_{2}+M_{3}.\label{pf:M1M2M3}
\end{eqnarray}
Since $L_{i}=\left(\frac{T_{i}}{\pi}-1\right)\left(e_{i}(1)+\frac{\pi}{1-\pi}e_{i}(0)\right)$ by (\ref{pf:T}), we have $M_{1}=\mathcal{L}$.

For $M_{2}+M_{3}$, using the relation 
\[
\Sigma_{t}^{-1}=\Sigma^{-1}-\Sigma^{-1}(\Sigma_{t}-\Sigma)\Sigma^{-1}+\Sigma^{-1}(\Sigma_{t}-\Sigma)^{\prime}\Sigma^{-1}(\Sigma_{t}-\Sigma)\Sigma_{t}^{-1},
\]
for $t=0,1$, we decompose 
\begin{eqnarray*}
M_{2} & = & -\left\{ \frac{1}{n}\sum_{i=1}^{n}\left(\frac{T_{i}}{\pi}-1\right)z_{i}^{\prime}\right\} \Sigma^{-1}\left\{ \frac{1}{n}\sum_{i=1}^{n}\left(\frac{T_{i}}{\pi}-1\right)z_{i}e_{i}(1)\right\} \\
 &  & +\left\{ \frac{1}{n}\sum_{i=1}^{n}\left(\frac{T_{i}}{\pi}-1\right)z_{i}^{\prime}\right\} \Sigma^{-1}(\Sigma_{1}-\Sigma)\Sigma^{-1}\left\{ \frac{1}{n}\sum_{i=1}^{n}\left(\frac{T_{i}}{\pi}-1\right)z_{i}e_{i}(1)\right\} \\
 &  & -\left\{ \frac{1}{n}\sum_{i=1}^{n}\left(\frac{T_{i}}{\pi}-1\right)z_{i}^{\prime}\right\} \Sigma^{-1}(\Sigma_{1}-\Sigma)^{\prime}\Sigma^{-1}(\Sigma_{1}-\Sigma)\Sigma_{1}^{-1}\left\{ \frac{1}{n}\sum_{i=1}^{n}\left(\frac{T_{i}}{\pi}-1\right)z_{i}e_{i}(1)\right\} \\
 & =: & M_{21}+M_{22}+M_{23},
\end{eqnarray*}
and 
\begin{eqnarray*}
M_{3} & = & \left(\frac{\pi}{1-\pi}\right)^{2}\left\{ \frac{1}{n}\sum_{i=1}^{n}\left(\frac{T_{i}}{\pi}-1\right)z_{i}^{\prime}\right\} \Sigma^{-1}\left\{ \frac{1}{n}\sum_{i=1}^{n}\left(\frac{T_{i}}{\pi}-1\right)z_{i}e_{i}(0)\right\} \\
 &  & -\left(\frac{\pi}{1-\pi}\right)^{2}\left\{ \frac{1}{n}\sum_{i=1}^{n}\left(\frac{T_{i}}{\pi}-1\right)z_{i}^{\prime}\right\} \Sigma^{-1}(\Sigma_{0}-\Sigma)\Sigma^{-1}\left\{ \frac{1}{n}\sum_{i=1}^{n}\left(\frac{T_{i}}{\pi}-1\right)z_{i}e_{i}(0)\right\} \\
 &  & +\left(\frac{\pi}{1-\pi}\right)^{2}\left\{ \frac{1}{n}\sum_{i=1}^{n}\left(\frac{T_{i}}{\pi}-1\right)z_{i}^{\prime}\right\} \Sigma^{-1}(\Sigma_{0}-\Sigma)^{\prime}\Sigma^{-1}(\Sigma_{0}-\Sigma)\Sigma_{0}^{-1}\left\{ \frac{1}{n}\sum_{i=1}^{n}\left(\frac{T_{i}}{\pi}-1\right)z_{i}e_{i}(0)\right\} \\
 & =: & M_{31}+M_{32}+M_{33}.
\end{eqnarray*}

{ 
The term $M_{21}+M_{31}$ can be written as 
\begin{equation}
M_{21}+M_{31}=B^{\mathrm{adj}}+\frac{1}{n}\sum_{i=1}^n \left\{\left(\frac{T_i}{\pi}-1\right)^2 P_{ii} \rho_i-B^{\mathrm{adj}}\right\}+\mathcal{W},\label{pf:M21}
\end{equation}
Lemma \ref{lem:T} implies
\begin{align} 
&{\mathbb E}\left[\frac{1}{n}\sum_{i=1}^n \left(\frac{T_i}{\pi}-1\right)^2 P_{ii} \rho_i\right]=B^{\mathrm{adj}} \sim \frac{1}{n}\sum_{i=1}^n P_{ii}\rho_i \le \sqrt{\frac{1}{n}\sum_{i=1}^n P_{ii}^2}\sqrt{\frac{1}{n}\sum_{i=1}^n \rho_{i}^2} \le n^{-1/2}(\mathcal{E}_{2}\kappa p)^{1/2}, \nonumber\\
&{\mathbb V}\left[\frac{1}{n}\sum_{i=1}^n \left(\frac{T_i}{\pi}-1\right)^2 P_{ii} \rho_i\right]
\sim\frac{1}{n^{2}}\sum_{i=1}^{n}P_{ii}^{2}\rho_{i}^{2} \le \frac{\mathcal{E}_{2}\kappa^{2}}{n},
\label{pf:B_rem}
\end{align} 
where the inequality follows from (\ref{pf:P}). Combining (\ref{pf:M1M2M3}), (\ref{pf:M21}) and (\ref{pf:B_rem}), we have 
\[
\hat{\tau}^{\mathrm{adj}}-\tau=B^{\mathrm{adj}}+\mathcal{L}+\mathcal{W}+(M_{22}+M_{32})+(M_{23}+M_{33})+O_p(n^{-1/2}({\mathcal E}_2\kappa^2)^{1/2}).
\]}
Thus, it is sufficient for (\ref{eq:sto}) to show 
\begin{equation}
(M_{22}+M_{32})+(M_{23}+M_{33})=O_{p}(n^{-1/2}\{\mathcal{E}_{2}(\kappa^{2}p+\kappa^{3}p(\log p)^{2})\}^{1/2}).\label{pf:M}
\end{equation}

We first consider the term $(M_{22}+M_{32})$. Since $\Sigma_{0}-\Sigma=-\frac{\pi}{1-\pi}(\Sigma_{1}-\Sigma)$,
we have 
\[
M_{22}+M_{32}=\left\{ \frac{1}{n}\sum_{i=1}^{n}\left(\frac{T_{i}}{\pi}-1\right)z_{i}^{\prime}\right\} \Sigma^{-1}(\Sigma_{1}-\Sigma)\Sigma^{-1}\left\{ \frac{1}{n}\sum_{i=1}^{n}\left(\frac{T_{i}}{\pi}-1\right)z_{i}\left(e_{i}(1)+\left(\frac{\pi}{1-\pi}\right)^{3}e_{i}(0)\right)\right\} .
\]
We obtain 
\begin{eqnarray*}
 &  & \mathbb{E}[(M_{22}+M_{32})^{2}]\\
 & \sim & \mathbb{E}\left[\frac{1}{n^{2}}\sum_{i,j,k,l,m,q}^{n}\left(\frac{T_{i}}{\pi}-1\right)\left(\frac{T_{j}}{\pi}-1\right)\left(\frac{T_{k}}{\pi}-1\right)\left(\frac{T_{l}}{\pi}-1\right)\left(\frac{T_{m}}{\pi}-1\right)\left(\frac{T_{q}}{\pi}-1\right)P_{ij}P_{jk}P_{lm}P_{mq}r_{k}r_{q}\right]\\
 & \sim & \frac{1}{n^{2}}\sum_{\substack{i,j,k\\
i\neq j\neq k
}
}\left(P_{ii}P_{ij}P_{jk}P_{kk}r_{j}r_{k}+P_{ii}P_{ij}P_{kj}P_{jk}r_{j}r_{k}+P_{ii}P_{ij}P_{kk}P_{kj}r_{j}^{2}+P_{ij}P_{ji}P_{jk}P_{kk}r_{i}r_{k}\right.\\
 &  & \qquad\qquad+P_{ij}P_{ji}P_{kj}P_{jk}r_{i}r_{k}+P_{ij}P_{ji}P_{kk}P_{kj}r_{i}r_{j}+P_{ij}P_{jj}P_{ik}P_{kk}r_{j}r_{k}+P_{ij}P_{jk}P_{ij}P_{jk}r_{k}^{2}\\
 &  & \qquad\qquad+P_{ij}P_{jk}P_{ik}P_{kj}r_{k}r_{j}+P_{ij}P_{jj}P_{ki}P_{ik}r_{j}r_{k}+P_{ij}P_{jk}P_{ji}P_{ik}r_{k}^{2}+P_{ij}P_{jk}P_{ki}P_{ij}r_{k}r_{j}\\
 &  & \left.\qquad\qquad+P_{ij}P_{jj}P_{kk}P_{ki}r_{j}r_{i}+P_{ij}P_{jk}P_{jk}P_{ki}r_{k}r_{i}+P_{ij}P_{jk}P_{kj}P_{ji}r_{k}r_{i}\right)\\
 & =: & A_{1}+\cdots+A_{15}.
\end{eqnarray*}

To calculate the orders of these terms, we will use the following inequalities
\begin{equation}
\frac{1}{n}\sum_{i,j,k}P_{ij}^{2}P_{kk}^{2}r_{j}^{2}=\frac{1}{n}\sum_{j,k}P_{jj}P_{kk}^{2}r_{j}^{2}\le\frac{\kappa}{n}\sum_{j,k}P_{jj}P_{kk}r_{j}^{2}=\frac{\kappa p}{n}\sum_{j}P_{jj}r_{j}^{2}\le\mathcal{E}_{2}\kappa^{2}p,\label{lem:A1}
\end{equation}
where the equalities follow from (\ref{pf:P}), and the inequalities use the definitions of $\kappa$ and $\mathcal{E}_{2}$.

\begin{equation}
\frac{1}{n}\sum_{i,j,k}P_{ij}^{2}P_{jk}^{2}r_{k}^{2}=\frac{1}{n}\sum_{j,k}P_{jj}P_{jk}^{2}r_{k}^{2}\le\frac{\kappa}{n}\sum_{j,k}P_{jk}^{2}r_{k}^{2}=\frac{\kappa}{n}\sum_{k}P_{kk}r_{k}^{2}\le\mathcal{E}_{2}\kappa^{2},\label{lem:A2}
\end{equation}
where the equalities follow from (\ref{pf:P}), and the inequalities use the definitions of $\kappa$ and $\mathcal{E}_{2}$.
\begin{eqnarray}
\frac{1}{n}\sum_{i,j,k}P_{ii}P_{ij}P_{jk}P_{kk}r_{j}^{2} & = & \frac{1}{n}\sum_{j}r_{j}^{2}\left(\sum_{i}P_{ii}P_{ij}\right)^{2}\le\frac{1}{n}\sum_{j}r_{j}^{2}\left(\sum_{i}P_{ii}^{2}\right)\left(\sum_{i}P_{ij}^{2}\right)\nonumber \\
 & \le & \frac{\kappa}{n}\sum_{j}r_{j}^{2}\left(\sum_{i}P_{ii}\right)P_{jj}\le\mathcal{E}_{2}\kappa^{2}p,\label{lem:A3}
\end{eqnarray}
where the first inequality follows from the Cauchy-Schwarz inequality, the second equality follows from the definition of $\kappa$ and (\ref{pf:P}), and the third inequality follows from the definitions of $\kappa$ and $\mathcal{E}_{2}$ and (\ref{pf:P}).
\begin{equation}
\frac{1}{n}\sum_{i,j,k}P_{ij}^{2}P_{kk}^{2}r_{k}^{2}=\frac{p}{n}\sum_{k}P_{kk}^{2}r_{k}^{2}\le\mathcal{E}_{2}\kappa^{2}p,\label{lem:A4}
\end{equation}
where the equality follows from (\ref{pf:P}), and the inequality uses the definitions of $\kappa$ and $\mathcal{E}_{2}$.
\begin{equation}
\frac{1}{n}\sum_{i,j,k}P_{ij}^{2}P_{jk}^{2}r_{i}^{2}=\frac{1}{n}\sum_{i}r_{i}^{2}\sum_{j}P_{ij}^{2}P_{jj}\le\frac{\kappa}{n}\sum_{i}r_{i}^{2}\sum_{j}P_{ij}^{2}=\frac{\kappa}{n}\sum_{i}r_{i}^{2}P_{ii}\le\mathcal{E}_{2}\kappa^{2},\label{lem:A5}
\end{equation}
where the equalities follow from (\ref{pf:P}), and the inequalities use the definitions of $\kappa$ and $\mathcal{E}_{2}$.
\begin{equation}
\frac{1}{n}\sum_{i,j,k}P_{ij}^{2}P_{kj}^{2}r_{j}^{2}=\frac{1}{n}\sum_{j}P_{jj}^{2}r_{j}^{2}\le\mathcal{E}_{2}\kappa^{2},\label{lem:A6}
\end{equation}
where the equality follows from (\ref{pf:P}), and the inequality uses the definitions of $\kappa$ and $\mathcal{E}_{2}$.
\begin{equation}
\frac{1}{n}\sum_{j,k}P_{jj}^{2}P_{kk}^{2}r_{j}^{2}\le\kappa^{3}\left(\sum_{k}P_{kk}\right)\frac{1}{n}\sum_{j}r_{j}^{2}\le\mathcal{E}_{2}\kappa^{3}p,\label{lem:A7-1}
\end{equation}
where the inequalities follow from the definitions of $\kappa$ and $\mathcal{E}_{2}$ and (\ref{pf:P}).
\begin{equation}
\frac{1}{n}\sum_{j,k}P_{jk}^{2}r_{k}^{2}=\frac{1}{n}\sum_{k}P_{kk}r_{k}^{2}\le\mathcal{E}_{2}\kappa,\label{lem:A7-2}
\end{equation}
where the equality follows from (\ref{pf:P}) and the inequality follows from the definitions of $\kappa$ and $\mathcal{E}_{2}$.

\begin{equation}
\frac{1}{n}\sum_{j,k}P_{jk}^{4}r_{k}^{2}\le(\max_{j,k}P_{jk}^{2})\frac{1}{n}\sum_{j,k}P_{jk}^{2}r_{k}^{2}=(\max_{j,k}P_{jk}^{2})\frac{1}{n}\sum_{k}P_{kk}r_{k}^{2}\le\mathcal{E}_{2}\kappa^{3},\label{lem:A9}
\end{equation}
where the equality follows from (\ref{pf:P}) and the inequality follows
from the definition of $\mathcal{E}_{2}$ and $\kappa$ and $|P_{jk}|\le P_{jj}$.

\begin{equation}
\frac{1}{n^{2}}\sum_{i,j,k}P_{ij}^{2}P_{jk}P_{ik}r_{k}^{2}\le\frac{1}{n^{2}}\sqrt{\sum_{i,j,k}P_{ij}^{2}P_{jk}^{2}r_{k}^{2}}\sqrt{\sum_{i,j,k}P_{ij}^{2}P_{ik}^{2}r_{k}^{2}}\le\frac{\mathcal{E}_{2}\kappa^{2}}{n},\label{lem:A-11}
\end{equation}
where the first inequality follows from the Cauchy-Schwarz inequality, and the second inequality follows from (\ref{lem:A2}).

Based on these inequalities, we can obtain the orders of $A_{1}\cdots,A_{15}$.
For $A_{1}$, 
\[
A_{1}=\frac{1}{n^{2}}\sum_{\substack{i,j,k\\
i\neq j\neq k
}
}P_{ii}P_{ij}P_{jk}P_{kk}r_{j}r_{k}\le\frac{1}{n^{2}}\sqrt{\sum_{i,j,k}P_{ij}^{2}P_{kk}^{2}r_{j}^{2}}\sqrt{\sum_{i,j,k}P_{ii}^{2}P_{jk}^{2}r_{k}^{2}}=\frac{1}{n}\left(\frac{1}{n}\sum_{i,j,k}P_{ij}^{2}P_{kk}^{2}r_{j}^{2}\right)=O\left(\frac{\mathcal{E}_{2}\kappa^{2}p}{n}\right),
\]
where the inequality follows from the Cauchy-Schwarz inequality, and the second equality follows from (\ref{lem:A1}). For $A_{2}$, 
\[
A_{2}=\frac{1}{n^{2}}\sum_{\substack{i,j,k\\
i\neq j\neq k
}
}P_{ii}P_{ij}P_{jk}^{2}r_{j}r_{k}\le\frac{1}{n}\sqrt{\frac{1}{n}\sum_{i,j,k}P_{ii}^{2}P_{jk}^{2}r_{j}^{2}}\sqrt{\frac{1}{n}\sum_{i,j,k}P_{ij}^{2}P_{jk}^{2}r_{k}^{2}}=O\left(\frac{\mathcal{E}_{2}\kappa^{2}p^{1/2}}{n}\right),
\]
where the inequality follows from the Cauchy-Schwarz inequality, and the second equality follows from (\ref{lem:A1}) and (\ref{lem:A2}). For $A_{3}$, 
\[
A_{3}=\frac{1}{n^{2}}\sum_{\substack{i,j,k\\
i\neq j\neq k
}
}P_{ii}P_{ij}P_{kk}P_{kj}r_{j}^{2}=O\left(\frac{\mathcal{E}_{2}\kappa^{2}p}{n}\right),
\]
where the second equality follows from (\ref{lem:A3}). For $A_{4}$,
\[
A_{4}=\frac{1}{n^{2}}\sum_{\substack{i,j,k\\
i\neq j\neq k
}
}P_{ij}^{2}P_{jk}P_{kk}r_{i}r_{k}\le\frac{1}{n}\sqrt{\frac{1}{n}\sum_{i,j,k}P_{ij}^{2}P_{jk}^{2}r_{i}^{2}}\sqrt{\frac{1}{n}\sum_{i,j,k}P_{ij}^{2}P_{kk}^{2}r_{k}^{2}}=O\left(\frac{\mathcal{E}_{2}\kappa^{2}p^{1/2}}{n}\right),
\]
where the inequality follows from the Cauchy-Schwarz inequality, and the second equality follows from (\ref{lem:A2}) and (\ref{lem:A4}). For $A_{5}$, 
\[
A_{5}=\frac{1}{n^{2}}\sum_{\substack{i,j,k\\
i\neq j\neq k
}
}P_{ij}^{2}P_{jk}^{2}r_{i}r_{k}\le\frac{1}{n^{2}}\sum_{i,j,k}P_{ij}^{2}P_{jk}^{2}(r_{i}^{2}+r_{k}^{2})=\frac{2}{n^{2}}\sum_{i,j,k}P_{ij}^{2}P_{jk}^{2}r_{i}^{2}=O\left(\frac{\mathcal{E}_{2}\kappa^{2}}{n}\right),
\]
where the third equality follows from (\ref{lem:A5}). For $A_{6}$,
\[
A_{6}=\frac{1}{n^{2}}\sum_{\substack{i,j,k\\
i\neq j\neq k
}
}P_{ij}^{2}P_{kk}P_{kj}r_{i}r_{j}\le\frac{1}{n}\sqrt{\frac{1}{n}\sum_{i,j,k}P_{ij}^{2}P_{kk}^{2}r_{i}^{2}}\sqrt{\frac{1}{n}\sum_{i,j,k}P_{ij}^{2}P_{kj}^{2}r_{j}^{2}}=O\left(\frac{\mathcal{E}_{2}\kappa^{2}p^{1/2}}{n}\right),
\]
where the second equality follows form (\ref{lem:A1}) and (\ref{lem:A6}).
For $A_{7}$, 
\begin{eqnarray*}
A_{7} & = & \frac{1}{n^{2}}\sum_{\substack{i,j,k\\
i\neq j\neq k
}
}P_{ij}P_{jj}P_{ik}P_{kk}r_{j}r_{k}\sim\frac{1}{n^{2}}\sum_{j,k}P_{jj}P_{jk}P_{kk}r_{j}r_{k}\\
 & \le & \frac{1}{n}\sqrt{\frac{1}{n}\sum_{j,k}P_{jj}^{2}P_{kk}^{2}r_{j}^{2}}\sqrt{\frac{1}{n}\sum_{j,k}P_{jk}^{2}r_{k}^{2}}=O\left(\frac{\mathcal{E}_{2}\kappa^{2}p^{1/2}}{n}\right),
\end{eqnarray*}
where the wave equality and the second equality follow from (\ref{pf:P}), the inequality follows from the Cauchy-Schwarz inequality, and the second equality follows from (\ref{lem:A7-1}) and (\ref{lem:A7-2}). For $A_{8}$, 
\[
A_{8}=\frac{1}{n^{2}}\sum_{\substack{i,j,k\\
i\neq j\neq k
}
}P_{ij}^{2}P_{jk}^{2}r_{k}^{2}\le\frac{1}{n^{2}}\sum_{i,j,k}P_{ij}^{2}P_{jk}^{2}r_{k}^{2}=O\left(\frac{\mathcal{E}_{2}\kappa^{2}}{n}\right),
\]
where the second equality follows from (\ref{lem:A2}). For $A_{9}$,
\[
A_{9}=\frac{1}{n^{2}}\sum_{\substack{i,j,k\\
i\neq j\neq k
}
}P_{ij}P_{jk}^{2}P_{ik}r_{k}r_{j}\sim\frac{1}{n^{2}}\sum_{j,k}P_{jk}^{3}r_{k}r_{j}\le\frac{1}{n}\sqrt{\frac{1}{n}\sum_{j,k}P_{jk}^{2}r_{k}^{2}}\sqrt{\frac{1}{n}\sum_{j,k}P_{jk}^{4}r_{j}^{2}}=O\left(\frac{\mathcal{E}_{2}\kappa^{2}}{n}\right),
\]
where the wave equality follows from (\ref{pf:P}), and the second equality follows from (\ref{lem:A7-2}) and (\ref{lem:A9}). For $A_{10}$,
\[
A_{10}=\frac{1}{n^{2}}\sum_{\substack{i,j,k\\
i\neq j\neq k
}
}P_{ij}P_{jj}P_{ik}^{2}r_{j}r_{k}\le\frac{1}{n}\sqrt{\sum_{i,j,k}P_{ij}^{2}P_{ik}^{2}r_{j}^{2}}\sqrt{\sum_{i,j,k}P_{jj}^{2}P_{ik}^{2}r_{k}^{2}}=O\left(\frac{\mathcal{E}_{2}\kappa^{2}p^{1/2}}{n}\right),
\]
where the inequality follows from the Cauchy-Schwarz inequality, and the second equality follows from (\ref{lem:A2}) and (\ref{lem:A1}). For $A_{11}$, 
\[
A_{11}=\frac{1}{n^{2}}\sum_{\substack{i,j,k\\
i\neq j\neq k
}
}P_{ij}^{2}P_{jk}P_{ik}r_{k}^{2}\sim\frac{1}{n^{2}}\sum_{i,j,k}P_{ij}^{2}P_{jk}P_{ik}r_{k}^{2}=O\left(\frac{\mathcal{E}_{2}\kappa^{2}}{n}\right),
\]
where the second equality follows from (\ref{lem:A-11}). For $A_{12}$,
\[
A_{12}=\frac{1}{n^{2}}\sum_{\substack{i,j,k\\
i\neq j\neq k
}
}P_{ij}^{2}P_{jk}P_{ik}r_{k}r_{j}\le\frac{1}{n}\sqrt{\sum_{i,j,k}P_{ij}^{2}P_{ik}^{2}r_{j}^{2}}\sqrt{\sum_{i,j,k}P_{ij}^{2}P_{jk}^{2}r_{k}^{2}}=O\left(\frac{\mathcal{E}_{2}\kappa^{2}}{n}\right),
\]
where the inequality follows from the Cauchy-Schwarz inequality, and the second equality follows from (\ref{lem:A2}). For $A_{13}$,
\[
A_{13}=\frac{1}{n^{2}}\sum_{\substack{i,j,k\\
i\neq j\neq k
}
}P_{ij}P_{jj}P_{kk}P_{ki}r_{j}r_{i}=O\left(\frac{\mathcal{E}_{2}\kappa^{2}p}{n}\right),
\]
since $A_{13}=A_{1}$. For $A_{14}$, 
\[
A_{14}=\frac{1}{n^{2}}\sum_{\substack{i,j,k\\
i\neq j\neq k
}
}P_{ij}P_{jk}^{2}P_{ki}r_{k}r_{i}\le\frac{1}{n}\sqrt{\sum_{i,j,k}P_{ij}^{2}P_{jk}^{2}r_{k}^{2}}\sqrt{\sum_{i,j,k}P_{jk}^{2}P_{ki}^{2}r_{i}^{2}}=O\left(\frac{\mathcal{E}_{2}\kappa^{2}}{n}\right),
\]
where the inequality follows from the Cauchy-Schwarz inequality, and the second equality follows from (\ref{lem:A2}). For $A_{15}$,
\[
A_{15}=\frac{1}{n^{2}}\sum_{\substack{i,j,k\\
i\neq j\neq k
}
}P_{ij}^{2}P_{jk}^{2}r_{k}r_{i}\le\frac{1}{n}\sqrt{\sum_{i,j,k}P_{ij}^{2}P_{jk}^{2}r_{k}^{2}}\sqrt{\sum_{i,j,k}P_{ij}^{2}P_{jk}^{2}r_{i}^{2}}=O\left(\frac{\mathcal{E}_{2}\kappa^{2}}{n}\right),
\]
where the inequality follows from the Cauchy-Schwarz inequality, and the second equality follows from (\ref{lem:A2}).

Combining these results, we have 
\[
M_{22}+M_{32}=O_{p}(n^{-1/2}(\mathcal{E}_{2}\kappa^{2}p)^{1/2}).
\]

It remains to check the orders of $M_{23}$ and $M_{32}$. The norm inequality implies 
\begin{eqnarray}
M_{23} & \le & \left\Vert \frac{1}{n}\sum_{i=1}^{n}\left(\frac{T_{i}}{\pi}-1\right)z_{i}\right\Vert _{2}\left\Vert \Sigma^{-1}\right\Vert _{\mathrm{op}}^{2}\left\Vert \Sigma_{1}^{-1}\right\Vert _{\mathrm{op}}\left\Vert \Sigma_{1}-\Sigma\right\Vert _{\mathrm{op}}^{2}\left\Vert \frac{1}{n}\sum_{i=1}^{n}\left(\frac{T_{i}}{\pi}-1\right)z_{i}e_{i}(1)\right\Vert _{2}\nonumber \\
 & = & O_{p}(n^{-1/2}(\mathcal{E}_{2}\kappa^{3}p(\log p)^{2})^{1/2}),\label{pf:M23}
\end{eqnarray}
where the equality follows from Lei and Ding (2021, Lemmas A.8 and A.9), i.e., 
\begin{align*}
 & \left\Vert \frac{1}{n}\sum_{i=1}^{n}\left(\frac{T_{i}}{\pi}-1\right)z_{i}\right\Vert _{2}=O_{p}(\sqrt{p/n}),\qquad\left\Vert \frac{1}{n}\sum_{i=1}^{n}\left(\frac{T_{i}}{\pi}-1\right)z_{i}e_{i}(1)\right\Vert _{2}=O_{p}(\sqrt{\mathcal{E}_{2}\kappa}),\\
 & \left\Vert \Sigma_{1}-\Sigma\right\Vert _{\mathrm{op}}=O_{p}(\sqrt{\kappa\log p}),\qquad\left\Vert \Sigma^{-1}\right\Vert _{\mathrm{op}}=O_{p}(1),\qquad\left\Vert \Sigma_{1}^{-1}\right\Vert _{\mathrm{op}}=O_{p}(1),
\end{align*}
that hold under our Assumption \ref{asm:1}. The same argument yields $M_{32}=O_{p}(n^{-1/2}(\mathcal{E}_{2}\kappa^{3}p(\log p)^{2})^{1/2})$. Therefore, we obtain (\ref{pf:M}) and the conclusion in (\ref{eq:sto}) follows for $\hat{\tau}^{\mathrm{adj}}$.

{ 
Next, we consider $\hat{\tau}^{\mathrm{bc}}$. By the expansion for $\hat{\tau}^{\mathrm{adj}}$, we have 
\[
\hat{\tau}^{\mathrm{bc}}-\tau=  \left(B^{\mathrm{adj}}+\frac{n_{0}}{n_{1}}\hat{\Delta}_{1}-\frac{n_{1}}{n_{0}}\hat{\Delta}_{0}\right)+\mathcal{L}+\mathcal{W} +O_{p}\left(n^{-1/2}(\mathcal{E}_{2}\kappa^{2}p)^{1/2}+n^{-1/2}(\mathcal{E}_{2}\kappa^{3}p(\log p)^{2})^{1/2}\right).
\]
Observe that 
\begin{align*}
\frac{n_{0}}{n_{1}}\hat{\Delta}_{1} & =  \frac{n_{0}}{n_{1}}\frac{1}{n}\sum_{i=1}^{n}\frac{T_{i}}{\pi}P_{ii}e_{i}(1)-\frac{n_{0}}{n_{1}}\frac{1}{n}\sum_{i=1}^{n}\frac{T_{i}}{\pi}P_{ii}z_{i}^{\prime}(\hat{\beta}_{1}-\beta_{1})\\
 & =  \frac{n_{0}}{n_{1}}\frac{1}{n}\sum_{i=1}^{n}P_{ii}e_{i}(1)+\frac{n_{0}}{n_{1}}\frac{1}{n}\sum_{i=1}^{n}\left(\frac{T_{i}}{\pi}-1\right)P_{ii}e_{i}(1)-\frac{n_{0}}{n_{1}}\left[\frac{1}{n}\sum_{i=1}^{n}\frac{T_{i}}{\pi}P_{ii}z_{i}^{\prime}\Sigma_{1}^{-1}\left\{\frac{1}{n}\sum_{i=1}^{n}\left(\frac{T_{i}}{\pi}-1\right)z_{i}e_{i}(1)\right\}\right]\\
  & :=  \frac{n_{0}}{n_{1}}\frac{1}{n}\sum_{i=1}^{n}P_{ii}e_{i}(1)+D_{11}+D_{12},
\end{align*}
where the first equality follows from the definitions of $\hat{\Delta}_{1}$ and $\hat{e}_{i}$, the second equality follows from the definitions of $\hat{\beta}_{1}$ and $\beta_{1}$ and (\ref{pf:e}). 
For $D_{11}$,
\begin{equation*}
{\mathbb E}[D_{11}^2] \lesssim \frac{1}{n^2}\sum_{i=1}^n P_{ii}^2 e_i(1)^2 =O\left(\frac{{\mathcal E}_2 \kappa^2}{n}\right),
\end{equation*}
and the Cauchy-Schwartz inequality implies $D_{11}=O_p\left(n^{-1/2}({\mathcal E}_2 \kappa^2)^{1/2}\right)$.
For $D_{12}$,
\begin{align*}
D_{12} &\lesssim \sqrt{\frac{1}{n}\sum_{i=1}^n P_{ii}^2}\sqrt{\left\{\frac{1}{n}\sum_{i=1}^{n}\left(\frac{T_{i}}{\pi}-1\right)z_{i}e_{i}(1)\right\}^{\prime}\Sigma_{1}^{-1}\left\{\frac{1}{n}\sum_{i=1}^{n}\left(\frac{T_{i}}{\pi}-1\right)z_{i}e_{i}(1)\right\}}\\
&=O_p\left(n^{-1/2}(\mathcal{E}_{2}\kappa^{2}p)^{1/2}\right).
\end{align*}
Thus
\begin{equation*}
\frac{n_{0}}{n_{1}}\hat{\Delta}_{1} 
 = \frac{n_{0}}{n_{1}}\frac{1}{n}\sum_{i=1}^{n}P_{ii}e_{i}(1)+O_p\left(n^{-1/2}({\mathcal E}_2 \kappa^2)^{1/2}+n^{-1/2}(\mathcal{E}_{2}\kappa^{2}p)^{1/2}\right).
\end{equation*}
A similar argument yields 
\[
\frac{n_{1}}{n_{0}}\hat{\Delta}_{0}=\frac{n_{1}}{n_{0}}\frac{1}{n}\sum_{i=1}^{n}P_{ii}e_{i}(0)+O_p\left(n^{-1/2}({\mathcal E}_2 \kappa^2)^{1/2}+n^{-1/2}(\mathcal{E}_{2}\kappa^{2}p)^{1/2}\right),
\]
Noting that $B_{adj}=-\frac{n_{0}}{n_{1}}\frac{1}{n}\sum_{i=1}^{n}P_{ii}e_{i}(1)+\frac{n_{1}}{n_{0}}\frac{1}{n}\sum_{i=1}^{n}P_{ii}e_{i}(0)$, we obtain the expansion in (\ref{eq:sto}) for $\hat{\tau}^{\mathrm{bc}}$.
}

Finally, let us consider $\hat{\tau}^{\mathrm{cf}}$. Observe that
\begin{eqnarray*}
\hat{\tau}^{\mathrm{cf}}-\tau & = & \frac{1}{n}\sum_{i=1}^{n}\left(\frac{T_{i}}{\pi}-1\right)\{e_{i}(1)-z_{i}^{\prime}(\hat{\beta}_{1}^{(i)}-\beta_{1})\}-\frac{1}{n}\sum_{i=1}^{n}\left(\frac{1-T_{i}}{1-\pi}-1\right)\{e_{i}(0)-z_{i}^{\prime}(\hat{\beta}_{0}^{(i)}-\beta_{0})\}\\
 & = & \frac{1}{n}\sum_{i=1}^{n}\left(\frac{T_{i}}{\pi}-1\right)\left(e_{i}(1)+\frac{\pi}{1-\pi}e_{i}(0)\right) \\
  & & - \frac{1}{n}\sum_{i=1}^{n}\left(\frac{T_{i}}{\pi}-1\right)z_{i}^{\prime}(\Sigma_{1}^{(i)})^{-1}\left\{ \frac{1}{n}\sum_{j\neq i}^{n}\left(\frac{T_{j}}{\pi}-1\right)z_{j}e_{j}(1)-\frac{1}{n}z_{i}e_{i}(1)\right\} \\
 & & +\left(\frac{\pi}{1-\pi}\right)^{2}\frac{1}{n}\sum_{i=1}^{n}\left(\frac{T_{i}}{\pi}-1\right)z_{i}^{\prime}(\Sigma_{0}^{(i)})^{-1}\left\{ \frac{1}{n}\sum_{j\neq i}^{n}\left(\frac{T_{i}}{\pi}-1\right)z_{i}e_{i}(0)-\frac{1}{n}z_{i}e_{i}(0)\right\} \\
 & =: & M_{1}+M_{2}^{\mathrm{cf}}+M_{3}^{\mathrm{cf}},
\end{eqnarray*}
where the first equality follows from the same argument in (\ref{pf:M1M2M3}), the second equality follows from the definition of $\hat{\beta}_{t}^{(i)}$, $y_{j}(t)=z_{j}^{\prime}\beta_{t}+e_{j}(t)$, and (\ref{pf:e}).

For $M_{2}^{\mathrm{cf}}$, using the relation 
\begin{equation}
(\Sigma_{t}^{(i)})^{-1}=\Sigma^{-1}-\Sigma^{-1}(\Sigma_{t}^{(i)}-\Sigma)\Sigma^{-1}+\Sigma^{-1}(\Sigma_{t}^{(i)}-\Sigma)\Sigma^{-1}(\Sigma_{t}^{(i)}-\Sigma)(\Sigma_{t}^{(i)})^{-1},\label{pf:sigi}
\end{equation}
for $t\in\{0,1\}$, we decompose 
\begin{eqnarray*}
M_{2}^{\mathrm{cf}} & = & -\frac{1}{n}\sum_{i=1}^{n}\left(\frac{T_{i}}{\pi}-1\right)z_{i}^{\prime}\Sigma^{-1}\left\{ \frac{1}{n}\sum_{j\neq i}^{n}\left(\frac{T_{j}}{\pi}-1\right)z_{j}e_{j}(1)-\frac{1}{n}z_{i}e_{i}(1)\right\} \\
 &  & +\frac{1}{n}\sum_{i=1}^{n}\left(\frac{T_{i}}{\pi}-1\right)z_{i}^{\prime}\Sigma^{-1}(\Sigma_{1}^{(i)}-\Sigma)\Sigma^{-1}\left\{ \frac{1}{n}\sum_{j\neq i}^{n}\left(\frac{T_{j}}{\pi}-1\right)z_{j}e_{j}(1)-\frac{1}{n}z_{i}e_{i}(1)\right\} \\
 &  & -\frac{1}{n}\sum_{i=1}^{n}\left(\frac{T_{i}}{\pi}-1\right)z_{i}^{\prime}\Sigma^{-1}(\Sigma_{1}^{(i)}-\Sigma)\Sigma^{-1}(\Sigma_{1}^{(i)}-\Sigma)(\Sigma_{1}^{(i)})^{-1}\left\{ \frac{1}{n}\sum_{j\neq i}^{n}\left(\frac{T_{j}}{\pi}-1\right)z_{j}e_{j}(1)-\frac{1}{n}z_{i}e_{i}(1)\right\} \\
 & =: & M_{21}^{\mathrm{cf}}+M_{22}^{\mathrm{cf}}+M_{23}^{\mathrm{cf}},
\end{eqnarray*}
and 
\begin{eqnarray*}
M_{3}^{\mathrm{cf}} & = & \left(\frac{\pi}{1-\pi}\right)^{2}\frac{1}{n}\sum_{i=1}^{n}\left(\frac{T_{i}}{\pi}-1\right)z_{i}^{\prime}\Sigma^{-1}\left\{ \frac{1}{n}\sum_{j\neq i}^{n}\left(\frac{T_{j}}{\pi}-1\right)z_{j}e_{j}(0)-\frac{1}{n}z_{i}e_{i}(0)\right\} \\
 &  & -\left(\frac{\pi}{1-\pi}\right)^{2}\frac{1}{n}\sum_{i=1}^{n}\left(\frac{T_{i}}{\pi}-1\right)z_{i}^{\prime}\Sigma^{-1}(\Sigma_{0}^{(i)}-\Sigma)\Sigma^{-1}\left\{ \frac{1}{n}\sum_{j\neq i}^{n}\left(\frac{T_{j}}{\pi}-1\right)z_{j}e_{j}(0)-\frac{1}{n}z_{i}e_{i}(0)\right\} \\
 &  & +\left(\frac{\pi}{1-\pi}\right)^{2}\frac{1}{n}\sum_{i=1}^{n}\left[\begin{array}{c}
\left(\frac{T_{i}}{\pi}-1\right)z_{i}^{\prime}\Sigma^{-1}(\Sigma_{0}^{(i)}-\Sigma)^{\prime}\Sigma^{-1}(\Sigma_{0}^{(i)}-\Sigma)(\Sigma_{0}^{(i)})^{-1}\\
\times\left\{ \frac{1}{n}\sum_{j\neq i}^{n}\left(\frac{T_{j}}{\pi}-1\right)z_{j}e_{j}(0)-\frac{1}{n}z_{i}e_{i}(0)\right\} 
\end{array}\right]\\
 & =: & M_{31}^{\mathrm{cf}}+M_{32}^{\mathrm{cf}}+M_{33}^{\mathrm{cf}}.
\end{eqnarray*}
The term $M_{21}^{\mathrm{cf}}+M_{31}^{\mathrm{cf}}$ can be written as 
\begin{eqnarray}
 &  & M_{21}^{\mathrm{cf}}+M_{31}^{\mathrm{cf}}\nonumber \\
 & = & -\frac{1}{n}\sum_{i=1}^{n}\left(\frac{T_{i}}{\pi}-1\right)z_{i}^{\prime}\Sigma^{-1}\left\{ \frac{1}{n}\sum_{j\neq i}^{n}\left(\frac{T_{j}}{\pi}-1\right)z_{j}e_{j}(1)\right\} +\frac{1}{n}\sum_{i=1}^{n}\left(\frac{T_{i}}{\pi}-1\right)P_{ii}e_{i}(1)\nonumber \\
 &  & +\left(\frac{\pi}{1-\pi}\right)^{2}\frac{1}{n}\sum_{i=1}^{n}\left(\frac{T_{i}}{\pi}-1\right)z_{i}^{\prime}\Sigma^{-1}\left\{ \frac{1}{n}\sum_{j\neq i}^{n}\left(\frac{T_{j}}{\pi}-1\right)z_{j}e_{j}(0)\right\} -\left(\frac{\pi}{1-\pi}\right)^{2}\frac{1}{n}\sum_{i=1}^{n}\left(\frac{T_{i}}{\pi}-1\right)P_{ii}e_{i}(0)\nonumber \\
 & = & \mathcal{W}+O_{p}(n^{-1/2}(\mathcal{E}_{2}\kappa^{2})^{1/2}),\label{pf:M21c}
\end{eqnarray}
where the second equality follows from the definition of $\mathcal{W}$ and, following Lemma \ref{lem:T}, 
\[
\mathbb{E}\left[\left(\frac{1}{n}\sum_{i=1}^{n}\left(\frac{T_{i}}{\pi}-1\right)P_{ii}\rho_{i}\right)^{2}\right]\sim\frac{1-\pi}{\pi}\frac{1}{n^{2}}\sum_{i=1}^{n}P_{ii}^{2}\rho_{i}^{2}=O(n^{-1}\mathcal{E}_{2}\kappa^{2}).
\]

Combining (\ref{pf:M21c}) and the bound from (\ref{pf:M1}) that we shall derive later, we have 
\[
\hat{\tau}^{\mathrm{cf}}-\tau = \mathcal{L} + \mathcal{W} + M_{22}^{\mathrm{cf}}+M_{23}^{\mathrm{cf}}+M_{32}^{\mathrm{cf}}+M_{33}^{\mathrm{cf}}.
\]
Thus, it is sufficient for the expansion of $\hat{\tau}^{\mathrm{cf}}$ in (\ref{eq:sto}) to show that 
\begin{equation}
M_{22}^{\mathrm{cf}}+M_{23}^{\mathrm{cf}}+M_{32}^{\mathrm{cf}}+M_{33}^{\mathrm{cf}}=O_{p}(n^{-1/2}\{\mathcal{E}_{2}(\kappa^{2}p^{1/2}+\kappa^{3}p(\log p)^{2})\}^{1/2}).\label{pf:Mc}
\end{equation}
For $M_{22}^{\mathrm{cf}}+M_{32}^{\mathrm{cf}}$, we have 
\begin{eqnarray*}
 &  & \mathbb{E}[(M_{22}^{\mathrm{cf}}+M_{32}^{\mathrm{cf}})^{2}]\\
 & \sim & \mathbb{E}\left[\frac{1}{n^{2}}\sum_{\substack{i,j,k,l,m,q\\
i\neq j,i\neq k,l\neq m,l\neq q
}
}^{n}\left(\frac{T_{i}}{\pi}-1\right)\left(\frac{T_{j}}{\pi}-1\right)\left(\frac{T_{k}}{\pi}-1\right)\left(\frac{T_{l}}{\pi}-1\right)\left(\frac{T_{m}}{\pi}-1\right)\left(\frac{T_{q}}{\pi}-1\right)P_{ij}P_{jk}P_{lm}P_{mq}r_{k}r_{q}\right]\\
 & \sim & A_{7}+A_{8}+A_{9}+A_{11}+A_{12}+A_{14}+A_{15}.
\end{eqnarray*}
Based on the orders obtained above, we have 
\[
M_{22}^{\mathrm{cf}}+M_{32}^{\mathrm{cf}}=O_{p}(n^{-1/2}(\mathcal{E}_{2}\kappa^{2}p^{1/2})^{1/2}).
\]
For $M_{23}^{\mathrm{cf}}$, we have 
\begin{eqnarray}
M_{23}^{\mathrm{cf}} & \lesssim & \left\Vert \frac{1}{n}\sum_{i=1}^{n}\left(\frac{T_{i}}{\pi}-1\right)z_{i}\right\Vert _{2}\left\Vert \Sigma^{-1}\right\Vert _{\mathrm{op}}^{2}\left\Vert \Sigma_{1}^{(i)}-\Sigma\right\Vert _{\mathrm{op}}^{2}\left\Vert (\Sigma_{1}^{(i)})^{-1}\right\Vert _{\mathrm{op}}\nonumber \\
 &  & \times\left\{ \left\Vert \frac{1}{n}\sum_{j\neq i}^{n}\left(\frac{T_{j}}{\pi}-1\right)z_{j}e_{j}(1)\right\Vert _{2}+\left\Vert \frac{1}{n}z_{i}e_{i}(1)\right\Vert _{2}\right\} \nonumber \\
 & = & O_{p}(n^{-1/2}(\mathcal{E}_{2}\kappa^{3}p(\log p)^{2})^{1/2}),\label{pf:M23cf}
\end{eqnarray}
where the wave inequality follows from Cauchy-Schwarz inequality and
repeated applications of (\ref{pf:sigi}), and the equality follows
from 
\begin{align*}
 & \left\Vert \frac{1}{n}\sum_{i=1}^{n}\left(\frac{T_{i}}{\pi}-1\right)z_{i}\right\Vert _{2}=O_{p}(\sqrt{p/n}),\qquad\left\Vert \frac{1}{n}\sum_{j\neq i}^{n}\left(\frac{T_{j}}{\pi}-1\right)z_{j}e_{j}(1)\right\Vert _{2}=O_{p}(\sqrt{\mathcal{E}_{2}\kappa}),\\
 & \left\Vert \frac{1}{n}z_{i}e_{i}(1)\right\Vert =O_{p}(n^{-1/2}(\mathcal{E}_{\infty}^{2}\kappa)^{1/2}),\qquad\left\Vert \Sigma_{1}^{(i)}-\Sigma\right\Vert _{\mathrm{op}}=O_{p}(\sqrt{\kappa\log p}),\qquad\left\Vert (\Sigma_{1}^{(i)})^{-1}\right\Vert _{\mathrm{op}}=O_{p}(1),
\end{align*}
under Assumption \ref{asm:1}.

\subsubsection{Proof of (ii)}

We firrst consider $\mathcal{L}$. Note that 
\[
\mathbb{E}[\mathcal{L}]=\mathbb{E}\left[\frac{T_{i}}{\pi}-1\right]\frac{1}{n}\sum_{i=1}^{n}\left(e_{i}(1)+\frac{\pi}{1-\pi}e_{i}(0)\right)=0,
\]
by Lemma \ref{lem:T}, and 
\begin{eqnarray}
\mathbb{V}(\mathcal{L}) & = & \frac{1}{n^{2}}\sum_{i=1}^{n}\mathbb{V}\left(\left(\frac{T_{i}}{\pi}-1\right)\left(e_{i}(1)+\frac{\pi}{1-\pi}e_{i}(0)\right)\right)\nonumber \\
 &  & +\frac{2}{n^{2}}\sum_{1\le i<j\le n}\mathbb{C}\left(\left(\frac{T_{i}}{\pi}-1\right)\left(e_{i}(1)+\frac{\pi}{1-\pi}e_{i}(0)\right),\left(\frac{T_{j}}{\pi}-1\right)\left(e_{j}(1)+\frac{\pi}{1-\pi}e_{j}(0)\right)\right)\nonumber \\
 & = & \frac{1-\pi}{\pi}\frac{1}{n^{2}}\sum_{i=1}^{n}\left(e_{i}(1)+\frac{\pi}{1-\pi}e_{i}(0)\right)^{2}\nonumber \\
 &  & -\frac{1-\pi}{\pi}\frac{2}{n^{2}(n-1)}\sum_{1\le i<j\le n}\left(e_{i}(1)+\frac{\pi}{1-\pi}e_{i}(0)\right)\left(e_{j}(1)+\frac{\pi}{1-\pi}e_{j}(0)\right)\nonumber \\
 & = & \frac{1-\pi}{\pi}\frac{1}{n}\left\{ \frac{1}{n-1}\sum_{i=1}^{n}\left(e_{i}(1)+\frac{\pi}{1-\pi}e_{i}(0)\right)^{2}\right\} \nonumber \\
 & = & \frac{n_{0}}{n_{1}}\frac{1}{n(n-1)}\sum_{i=1}^{n}e_{i}(1)^{2}+\frac{n_{1}}{n_{0}}\frac{1}{n(n-1)}\sum_{i=1}^{n}e_{i}(0)^{2}+\frac{2}{n(n-1)}\sum_{i=1}^{n}e_{i}(1)e_{i}(0)\nonumber \\
 & \lesssim & \frac{\mathcal{E}_{2}}{n},\label{pf:sigL}
\end{eqnarray}
where the second equality follows from Lemma \ref{lem:T}, the third equality follows from the relation $\frac{1}{n-1}\sum_{i=1}^{n}(a_{i}-\bar{a})^{2}=\frac{1}{n}\sum_{i=1}^{n}a_{i}^{2}-\frac{2}{n(n-1)}\sum_{i<j}a_{i}a_{j}$ and (\ref{pf:e}), and the wave inequality follows from the definition of $\mathcal{E}_{2}$ and Cauchy-Schwarz inequality. Thus, Chebyshev's inequality implies 
\begin{equation}
\mathcal{L}=O_{p}(n^{-1/2}\mathcal{E}_{2}^{1/2}).\label{pf:M1}
\end{equation}

Next, we consider $\mathcal{W}$. Using the fact that 
\[
\mathcal{W}=\frac{1}{2n^{2}}\sum_{i=1}^{n}\sum_{j\neq i}^{n}\left(\frac{T_{i}}{\pi}-1\right)\left(\frac{T_{j}}{\pi}-1\right)z_{i}^{\prime}\Sigma^{-1}z_{j}(\rho_{i}+\rho_{j})=\frac{1}{n}\sum_{i=1}^{n}\sum_{j\neq i}^{n}\left(\frac{T_{i}}{\pi}-1\right)\left(\frac{T_{j}}{\pi}-1\right)P_{ij}\rho_{j},
\]
we have 
\begin{eqnarray}
\mathbb{E}[\mathcal{W}] & = & -\frac{1}{n(n-1)}\frac{1-\pi}{\pi}\sum_{i=1}^{n}\sum_{j\neq i}^{n}P_{ij}\rho_{j}=\frac{1}{n(n-1)}\frac{1-\pi}{\pi}\sum_{i=1}^{n}P_{ii}\rho_{i}\nonumber \\
 & \le & \frac{1}{n^{1/2}(n-1)}\frac{1-\pi}{\pi}\sqrt{\sum_{i=1}^{n}P_{ii}^{2}}\sqrt{\frac{1}{n}\sum_{i=1}^{n}\rho_{i}^{2}}=O(n^{-3/2}(\mathcal{E}_{2}\kappa p)^{1/2}),\label{pf:EW}
\end{eqnarray}
where the first equality follows from Lemma \ref{lem:T}, the second equality follows from $\sum_{j=1}^{n}\rho_{j}z_{j}=0$ (by (\ref{pf:e})), the inequality follows from the Cauchy-Schwarz inequality, and the last equality follows from (\ref{pf:P}) and the definitions of $\mathcal{E}_{2}$ and $\kappa$. To characterize the order of $\mathbb{E}[\mathcal{W}^{2}]$, decompose 
\begin{eqnarray*}
 &  & \mathbb{E}[\mathcal{W}^{2}]\\
 & = & \mathbb{E}\left[\frac{1}{n^{2}}\sum_{\substack{i,j,k,l\\
i\neq j,k\neq l
}
}\left(\frac{T_{i}}{\pi}-1\right)\left(\frac{T_{j}}{\pi}-1\right)\left(\frac{T_{k}}{\pi}-1\right)\left(\frac{T_{l}}{\pi}-1\right)P_{ij}P_{kl}\rho_{j}\rho_{l}\right]\\
 & = & \frac{1}{n^{2}}\sum_{i=1}^{n}\sum_{j\neq i}^{n}\mathbb{E}\left[\left(\frac{T_{i}}{\pi}-1\right)^{2}\left(\frac{T_{j}}{\pi}-1\right)^{2}\right]P_{ij}^{2}(\rho_{i}\rho_{j}+\rho_{j}^{2})\\
 &  & +\frac{1}{n^{2}}\sum_{\substack{i,j,k\\
i\neq j\neq k
}
}\mathbb{E}\left[\left(\frac{T_{i}}{\pi}-1\right)^{2}\left(\frac{T_{j}}{\pi}-1\right)\left(\frac{T_{k}}{\pi}-1\right)\right]\left(P_{ij}P_{ik}\rho_{j}\rho_{k}+P_{ij}P_{ki}\rho_{j}\rho_{i}+P_{ij}P_{jk}\rho_{j}\rho_{k}+P_{ij}P_{kj}\rho_{j}^{2}\right)\\
 &  & +\frac{1}{n^{2}}\sum_{\substack{i,j,k,l\\
i\neq j\neq k\neq l
}
}\mathbb{E}\left[\left(\frac{T_{i}}{\pi}-1\right)\left(\frac{T_{j}}{\pi}-1\right)\left(\frac{T_{k}}{\pi}-1\right)\left(\frac{T_{l}}{\pi}-1\right)\right]P_{ij}P_{kl}\rho_{j}\rho_{l}\\
 & =: & V_{1}+V_{2}+V_{3}.
\end{eqnarray*}
For $V_{1}$, note that
\begin{eqnarray}
V_{1} & = & \frac{(1-\pi)^{2}}{\pi^{2}}\frac{1}{n(n-1)}\sum_{i=1}^{n}\sum_{j\neq i}^{n}P_{ij}^{2}(\rho_{i}\rho_{j}+\rho_{j}^{2})\le\frac{2(1-\pi)^{2}}{\pi^{2}}\frac{1}{n(n-1)}\sum_{i=1}^{n}\sum_{j\neq i}^{n}P_{ij}^{2}\rho_{j}^{2}\nonumber \\
 & = & \frac{2}{n^{2}}\sum_{i=1}^{n}(P_{ii}-P_{ii}^{2})\rho_{i}^{2}=O(n^{-1}\mathcal{E}_{2}\kappa)+O(n^{-1}\mathcal{E}_{2}\kappa^{2})=O(n^{-1}\mathcal{E}_{2}\kappa),\label{pf:V1}
\end{eqnarray}
where the first equality follows from Lemma \ref{lem:T}, the inequality follows from $\rho_{i}\rho_{j}\le\frac{1}{2}(\rho_{i}^{2}+\rho_{j}^{2})$, the second equality follows from (\ref{pf:P}), the third equality follows from the definitions of $\kappa$ and $\mathcal{E}_{2}$, and the last equality follows from $\kappa\le1$. For $V_{2}$, note that 
\begin{eqnarray*}
V_{2} & \sim & \frac{1}{n^{3}}\sum_{\substack{i,j,k\\
i\neq j\neq k
}
}P_{ij}P_{ik}(\rho_{i}+\rho_{j})(\rho_{i}+\rho_{k})\\
 & = & \frac{1}{n^{3}}\sum_{\substack{i,j\\
i\neq j
}
}P_{ij}(1-P_{ii}-P_{ij})(\rho_{i}+\rho_{j})\rho_{i}-\frac{1}{n^{3}}\sum_{\substack{i,j\\
i\neq j
}
}P_{ij}(P_{ii}\rho_{i}+P_{ij}\rho_{j})(\rho_{i}+\rho_{j})\\
 & \sim & \frac{1}{n^{3}}\sum_{\substack{i,j\\
i\neq j
}
}P_{ij}(\rho_{i}\rho_{j}+\rho_{j}^{2})=-\frac{1}{n^{3}}\sum_{j=1}^{n}P_{jj}\rho_{j}^{2}+\frac{1}{n^{3}}\sum_{j=1}^{n}(1-P_{jj})\rho_{j}^{2}\\
 & = & O(n^{-2}\mathcal{E}_{2}),
\end{eqnarray*}
where the first wave relation follows from Lemma \ref{lem:T}, the first equality follow from (\ref{pf:P}), the second wave relation follows from the similar inequalities as (\ref{lem:A1})-(\ref{lem:A-11}) (by replacing $r_{i}$ with $\rho_{i}$), the second equality follows from $\frac{1}{n^{2}}\sum_{i,j}P_{ij}\rho_{i}\rho_{j}=0$ and (\ref{pf:P}), and the last equality follows from the Cauchy-Schwarz inequality, $P_{jj}<1$ in (\ref{pf:P}), and the definition of $\mathcal{E}_{2}$. For $V_{3}$, note that 
\[
V_{3} \sim \frac{1}{n^{4}}\sum_{\substack{i,j,k,l\\
i\neq j\neq k\neq l}}P_{ij}P_{kl}\rho_{j}\rho_{l}\sim\frac{1}{n^{4}}\sum_{j,l}\rho_{j}\rho_{l} \le \frac{1}{2n^{4}}\sum_{j,l}(\rho_{j}^{2}+\rho_{l}^{2})=O(n^{-2}\mathcal{E}_{2}),
\]
where where the first wave relation follows from Lemma \ref{lem:T}, the second wave relation follows from (\ref{pf:P}), the inequality folllows from $\rho_{i}\rho_{j}\le\frac{1}{2}(\rho_{i}^{2}+\rho_{j}^{2})$, and the equality follows from the definition of $\mathcal{E}_{2}$. 

Combining these results, we obtain $\mathbb{E}[\mathcal{W}^{2}]=O(n^{-1}\mathcal{E}_{2}\kappa)$. Therefore, since (\ref{pf:EW}) implies $\mathbb{V}(\mathcal{W})\sim\mathbb{E}[\mathcal{W}^{2}]$, Chebyshev's inequality implies $\mathcal{W}=O_{p}(n^{-1/2}(\mathcal{E}_{2}\kappa)^{1/2})$.

We now prove the statements on $\sigma_{L}^{2}$ and $\sigma_{W}^{2}$. For $\sigma_{L}^{2}=\mathbb{V}(\sqrt{n}\mathcal{L})$, (\ref{pf:sigL}) implies $\sigma_{L}^{2}=O(\mathcal{E}_{2})$ and 
\begin{eqnarray*}
\sigma_{L}^{2} & = & \frac{n_{0}}{n_{1}}\frac{1}{n-1}\sum_{i=1}^{n}e_{i}(1)^{2}+\frac{n_{1}}{n_{0}}\frac{1}{n-1}\sum_{i=1}^{n}e_{i}(0)^{2}+\frac{2}{n-1}\sum_{i=1}^{n}e_{i}(1)e_{i}(0)\\
 & = & \frac{n}{n_{1}(n-1)}\sum_{i=1}^{n}e_{i}(1)^{2}+\frac{n}{n_{0}(n-1)}\sum_{i=1}^{n}e_{i}(0)^{2}-\frac{1}{n-1}\sum_{i=1}^{n}(e_{i}(1)-e_{i}(0))^{2},
\end{eqnarray*}
by direct algebra.

For $\sigma_{W}^{2}=\mathbb{V}(\sqrt{n}\mathcal{W})$, (\ref{pf:V1}) implies $\sigma_{W}^{2}=O_{p}(\mathcal{E}_{2}\kappa)$, and
\begin{eqnarray*}
\sigma_{W}^{2} & \sim & \frac{(1-\pi)^{2}}{\pi^{2}}\frac{1}{n-1}\sum_{i=1}^{n}\sum_{j\neq i}^{n}P_{ij}^{2}(\rho_{j}^{2}+\rho_{i}\rho_{j})\\
 & = & \frac{n_{0}^{2}}{n_{1}^{2}(n-1)}\sum_{j=1}^{n}(P_{ii}-P_{jj}^{2})e_{j}(1)^{2}+\frac{n_{1}^{2}}{n_{0}^{2}(n-1)}\sum_{j=1}^{n}(P_{jj}-P_{jj}^{2})e_{j}(0)^{2}-\frac{2}{n-1}\sum_{j=1}^{n}(P_{jj}-P_{jj}^{2})e_{j}(1)e_{j}(0)\\
 & & +\frac{n_{0}^{2}}{n_{1}^{2}(n-1)}\sum_{i=1}^{n}\sum_{j\neq i}^{n}P_{ij}^{2}\left\{ -e_{i}(1)+\left(\frac{\pi}{1-\pi}\right)^{2}e_{i}(0)\right\} \left\{ -e_{j}(1)+\left(\frac{\pi}{1-\pi}\right)^{2}e_{j}(0)\right\},
\end{eqnarray*}
where the equality follows from (\ref{pf:P}).


Finally, we check the order of $\mathbb{C}(\sqrt{n}\mathcal{L},\sqrt{n}\mathcal{W})$.
Let $\xi_{k}=e_{k}(1)+\frac{\pi}{1-\pi}e_{k}(0)$. Observe that 
\begin{eqnarray*}
 &  & \mathbb{C}(\sqrt{n}\mathcal{L},\sqrt{n}\mathcal{W})\\
 & = & \mathbb{C}\left(\frac{1}{\sqrt{n}}\sum_{k=1}^{n}\left(\frac{T_{k}}{\pi}-1\right)\xi_{k},\frac{1}{\sqrt{n}}\sum_{i=1}^{n}\sum_{j\neq i}^{n}\left(\frac{T_{i}}{\pi}-1\right)\left(\frac{T_{j}}{\pi}-1\right)P_{ij}\rho_{j}\right)\\
 & = & \frac{1}{n}\sum_{\substack{i,j,k\\
i\neq j
}
}\mathbb{E}\left[\left(\frac{T_{k}}{\pi}-1\right)\left(\frac{T_{i}}{\pi}-1\right)\left(\frac{T_{j}}{\pi}-1\right)\right]\xi_{k}P_{ij}\rho_{j}\\
 & = & \frac{1}{n}\sum_{\substack{i,j\\
i\neq j
}
}\mathbb{E}\left[\left(\frac{T_{i}}{\pi}-1\right)^{2}\left(\frac{T_{j}}{\pi}-1\right)\right]P_{ij}(\xi_{i}\rho_{j}+\xi_{j}\rho_{j}) +\frac{1}{n}\sum_{\substack{i,j,k\\
k\neq i\neq j
}
}\mathbb{E}\left[\left(\frac{T_{k}}{\pi}-1\right)\left(\frac{T_{i}}{\pi}-1\right)\left(\frac{T_{j}}{\pi}-1\right)\right]\xi_{k}P_{ij}\rho_{j}\\
 & =: & C_{1}+C_{2}.
\end{eqnarray*}
For $C_{1}$, 
\[
C_{1} \sim\frac{1}{n^{2}}\sum_{\substack{i,j\\
i\neq j}}P_{ij}(\xi_{i}\rho_{j}+\xi_{j}\rho_{j}) = -\frac{1}{n^{2}}\sum_{j}P_{jj}\xi_{j}\rho_{j}+\frac{1}{n^{2}}\sum_{j=1}^{n}(1-P_{jj})\xi_{j}\rho_{j} =O(n^{-1}\mathcal{E}_{2}),
\]
where the wave relation follows from Lemma \ref{lem:T}, the first equality follows from $\frac{1}{n^{2}}\sum_{i,j}P_{ij}\xi_{i}\rho_{j}=0$ and (\ref{pf:P}), and the second equality follows from the Cauchy-Schwarz inequality, $P_{jj}<1$ in (\ref{pf:P}), and the definition of $\mathcal{E}_{2}$. For $C_{2}$, 
\[
C_{2} \sim\frac{1}{n^{3}}\sum_{\substack{i,j,k\\
k\neq i\neq j}}\xi_{k}P_{ij}\rho_{j}\sim\frac{1}{n^{3}}\sum_{j,k}\xi_{k}\rho_{j} \le\frac{1}{2n^{3}}\sum_{j,k}(\xi_{k}^{2}+\rho_{j}^{2})=O(n^{-1}\mathcal{E}_{2}),
\]
where the first wave relation follows from Lemma \ref{lem:T}, the second wave relation follows from (\ref{pf:P}), the inequality follows from $\xi_{k}\rho_{j}\le\frac{1}{2}(\xi_{k}^{2}+\rho_{j}^{2})$, and the equality follows from the definition of $\mathcal{E}_{2}$. 

\subsection{Proof of Theorem \ref{thm:var1}}
{ 
We first show the statement on $\hat{\sigma}_{\text{HC0}}^{2}$. Decompose
\begin{equation*}
\hat{\sigma}_{\text{HC0}}^{2}  = \frac{1}{n_1-1}\sum_{i=1}^{n}\frac{T_{i}}{\pi}\hat{e}_i^{2}+\frac{1}{n_0-1}\sum_{i=1}^{n}\frac{1-T_{i}}{1-\pi}\hat{e}_i^{2}  =: \hat{\sigma}_{1,\text{HC0}}^{2}+\hat{\sigma}_{0,\text{HC0}}^{2}.
\end{equation*}
For $\hat{\sigma}_{1,\text{HC0}}^{2}$, we further decompose 
\begin{eqnarray*}
\hat{\sigma}_{1,\text{HC0}}^{2} & = & \frac{1}{n_1-1}\sum_{i=1}^{n}\frac{T_i}{\pi}\left\{e_i(1)-z_i^{\prime}(\hat{\beta}_1-\beta_1)\right\}^2\\
& = &  \frac{1}{n_{1}-1}\sum_{i=1}^{n}\frac{T_{i}}{\pi}e_{i}(1)^{2} -\frac{n}{n_{1}-1}\left(\frac{1}{n}\sum_{i=1}^{n}\frac{T_{i}}{\pi}z_{i}e_{i}(1)\right)^{\prime}\Sigma_{1}^{-1}\left(\frac{1}{n}\sum_{j=1}^{n}\frac{T_{j}}{\pi}z_{j}e_{j}(1)\right)\\
 & =: & \hat{\sigma}_{11,\text{HC0}}^{2}+\hat{\sigma}_{12,\text{HC0}}^{2}.
\end{eqnarray*}
For $\hat{\sigma}_{12,\text{HC0}}^{2}$, decompose
\begin{eqnarray*}
\hat{\sigma}_{12,\text{HC0}}^{2}  & = &  -\frac{n}{n_{1}-1}\left\{ \frac{1}{n}\sum_{i=1}^{n}\left(\frac{T_{i}}{\pi}-1\right)z_{i}e_{i}(1)\right\} ^{\prime}\Sigma^{-1}\left\{ \frac{1}{n}\sum_{j=1}^{n}\left(\frac{T_{j}}{\pi}-1\right)z_{j}e_{j}(1)\right\} \\
 & &  +\frac{n}{n_{1}-1}\left\{ \frac{1}{n}\sum_{i=1}^{n}\left(\frac{T_{i}}{\pi}-1\right)z_{i}e_{i}(1)\right\}^{\prime}\Sigma^{-1}(\Sigma_{1}-\Sigma)\Sigma_{1}^{-1}\left\{ \frac{1}{n}\sum_{j=1}^{n}\left(\frac{T_{j}}{\pi}-1\right)z_{j}e_{j}(1)\right\} \\
 & =: & \hat{\sigma}_{121,\text{HC0}}^{2}+\hat{\sigma}_{122,\text{HC0}}^{2},
\end{eqnarray*}
where the first equality follows from (\ref{pf:e}) and the relation $\Sigma_{1}^{-1}=\Sigma^{-1}-\Sigma^{-1}(\Sigma_{1}-\Sigma)\Sigma_{1}^{-1}$. By the same argument as in (\ref{pf:M23}), it holds $\hat{\sigma}_{122,\text{HC0}}^{2}=O_p\left(\mathcal{E}_{2}\kappa^{3/2}(\log p)^{1/2}\right)=o_{p}(\mathcal{E}_{2}\kappa)$.

For $\hat{\sigma}_{121,\text{HC0}}^{2}$, its mean is
\begin{align*}
{\mathbb E}[\hat{\sigma}_{121,\text{HC0}}^{2}]
&= -\frac{1}{n_1-1}\sum_{i=1}^n{\mathbb E}\left[\left(\frac{T_i}{\pi}-1\right)^2\right]P_{ii}e_i(1)^2
-\frac{1}{n_1-1}\sum_{i=1}^n\sum_{j\neq i}^n{\mathbb E}\left[\left(\frac{T_i}{\pi}-1\right)\left(\frac{T_j}{\pi}-1\right)\right]P_{ij}e_i(1)e_j(1)\\
&=-\frac{n_0}{(n_1-1)n_1}\sum_{i=1}^n P_{ii}e_i(1)^2+\frac{n_0}{n_1(n_1-1)(n-1)}\sum_{i=1}^n\sum_{j\neq i}^n P_{ij}e_i(1)e_j(1)\\
&=-\frac{n_0}{(n_1-1)n_1}\sum_{i=1}^n P_{ii}e_i(1)^2-\frac{n_0}{n_1(n_1-1)(n-1)}\sum_{i=1}^nP_{ii}e_i(1)^2\\
&=-\frac{n_0}{n_1^2}\sum_{i=1}^n P_{ii}e_i(1)^2+O\left(\frac{\mathcal{E}_2\kappa}{n}\right),
\end{align*}
where the third equality follows from Lemma \ref{lem:T}, (\ref{pf:e}) and (\ref{pf:P}), and the fourth equality follows from $P_{ii}\le\kappa$.
The variance of $\hat{\sigma}_{121,\text{HC0}}^{2}$ is 
\begin{eqnarray*}
& & {\mathbb V}[(\hat{\sigma}_{121,\text{HC0}}^{2})^2]\\
& = & {\mathbb E}\left[\frac{1}{(n_1-1)^2}\sum_{i,i_1,j,j_1=1}^n\left(\frac{T_{i}}{\pi}-1\right)\left(\frac{T_{i_1}}{\pi}-1\right)\left(\frac{T_{j}}{\pi}-1\right)\left(\frac{T_{j_1}}{\pi}-1\right)P_{ij}P_{i_1j_1}e_{i}(1)e_{i_1}(1)e_{j}(1)e_{j_1}(1)\right]\\
& & -({\mathbb E}[\hat{\sigma}_{121,\text{HC0}}^{2}])^2\\
&\lesssim & \frac{1}{n^2}\sum_{i,j} P_{ij}^2 e_i(1)^2e_j(1)^2 = O(({\mathcal E}_2\kappa^2)^2),
\end{eqnarray*}
where the wave inequality follows from Lemma \ref{lem:T} and the last equality follows from $P_{ij}^2 \le P_{ii}^2P_{jj}^2 \le \kappa^4$.
Thus, the Cauchy-Schwarz inequality implies that $\hat{\sigma}_{121,\text{HC0}}^{2}=-\frac{n_0}{n_1^2}\sum_{i=1}^n P_{ii}e_i(1)^2+o_p({\mathcal E}_2\kappa)$.
Combining these results, we obtain
\begin{equation}\label{pf:HC01}
\hat{\sigma}_{1,\text{HC0}}^{2}=\frac{n}{n_1(n-1)}\sum_{i=1}^n e_i(1)^2-\frac{n_0}{n_1^2}\sum_{i=1}^n P_{ii}e_i(1)^2+o_p({\mathcal E}_2\kappa).
\end{equation}
The same argument yields $\hat{\sigma}_{0,\text{HC0}}^{2}=\frac{n}{n_{0}(n-1)}\sum_{i=1}^{n}e_{i}(0)^{2}-\frac{n_{1}}{n_{0}^{2}}\sum_{i=1}^{n}P_{ii}e_{i}(0)^{2}+o_p({\mathcal E}_2\kappa)$, and the conclusion for $\hat{\sigma}_{\text{HC0}}^{2}$ follows.

We next show the statement on $\hat{\sigma}_{\text{HC3}}^{2}$. Decompose
\begin{equation*}
\hat{\sigma}_{\text{HC3}}^{2}  =  \frac{1}{n}\sum_{i=1}^{n}\left\{ \frac{T_{i}}{\pi}(Y_{i}-z_{i}^{\prime}\hat{\beta}_{1}^{(i)})\right\} ^{2}+\frac{1}{n}\sum_{i=1}^{n}\left\{ \frac{1-T_{i}}{1-\pi}(Y_{i}-z_{i}^{\prime}\hat{\beta}_{0}^{(i)})\right\} ^{2}
  =:  \hat{\sigma}_{1,\text{HC3}}^{2}+\hat{\sigma}_{0,\text{HC3}}^{2}.
\end{equation*}
For $\hat{\sigma}_{1,\text{HC3}}^{2}$, we further decompose 
\begin{eqnarray*}
\hat{\sigma}_{1,\text{HC3}}^{2} & = &  \frac{1}{n}\sum_{i=1}^{n}\left\{ \frac{T_{i}}{\pi}e_{i}(1)-\frac{T_{i}}{\pi}z_{i}^{\prime}(\hat{\beta}_{1}^{(i)}-\beta_{1})\right\} ^{2}\\
 & = &  \frac{1}{n_{1}}\sum_{i=1}^{n}\frac{T_{i}}{\pi}e_{i}(1)^{2}
 +\frac{1}{n_{1}}\sum_{i=1}^{n}\left(\frac{1}{n}\sum_{j\neq i}^{n}\frac{T_{j}}{\pi}z_{j}e_{j}(1)\right)^{\prime}(\Sigma_{1}^{(i)})^{-1}\frac{T_{i}}{\pi}z_{i}z_{i}^{\prime}(\Sigma_{1}^{(i)})^{-1}\left(\frac{1}{n}\sum_{j\neq i}^{n}\frac{T_{j}}{\pi}z_{j}e_{j}(1)\right)\\
 &  & -\frac{2n}{n_{1}}\left(\frac{1}{n}\sum_{i=1}^{n}\frac{T_{i}}{\pi}z_{i}e_{i}(1)\right)^{\prime}(\Sigma_{1}^{(i)})^{-1}\left(\frac{1}{n}\sum_{j\neq i}^{n}\frac{T_{j}}{\pi}z_{j}e_{j}(1)\right)\\
 & =: &  \hat{\sigma}_{11,\text{HC3}}^{2}+\hat{\sigma}_{12,\text{HC3}}^{2}+\hat{\sigma}_{13,\text{HC3}}^{2}.
\end{eqnarray*}
For $\hat{\sigma}_{13,\text{HC3}}^{2}$, decompose
\begin{eqnarray*}
\hat{\sigma}_{13,\text{HC3}}^{2} & = & -\frac{2n}{n_{1}}\left\{\frac{1}{n}\sum_{i=1}^{n}\left(\frac{T_{i}}{\pi}-1\right)z_{i}e_{i}(1)\right\}^{\prime}(\Sigma_{1}^{(i)})^{-1}\left\{\frac{1}{n}\sum_{j\neq i}^{n}\left(\frac{T_{j}}{\pi}-1\right)z_{j}e_{j}(1)-z_{i}e_{i}(1)\right\} \\
 & = & -\frac{2n}{n_{1}}\left\{ \frac{1}{n}\sum_{i=1}^{n}\left(\frac{T_{i}}{\pi}-1\right)z_{i}e_{i}(1)\right\} ^{\prime}\Sigma^{-1}\left\{ \frac{1}{n}\sum_{j\neq i}^{n}\left(\frac{T_{j}}{\pi}-1\right)z_{j}e_{j}(1)-\frac{1}{n}z_{i}e_{i}(1)\right\} \\
 &  & +\frac{2n}{n_{1}}\left\{ \frac{1}{n}\sum_{i=1}^{n}\left(\frac{T_{i}}{\pi}-1\right)z_{i}e_{i}(1)\right\} ^{\prime}\Sigma^{-1}(\Sigma_{1}^{(i)}-\Sigma)(\Sigma_{1}^{(i)})^{-1}\left\{ \frac{1}{n}\sum_{j\neq i}^{n}\left(\frac{T_{j}}{\pi}-1\right)z_{j}e_{j}(1)-\frac{1}{n}z_{i}e_{i}(1)\right\} \\
 & =: & \hat{\sigma}_{131,\text{HC3}}^{2}+\hat{\sigma}_{132,\text{HC3}}^{2},
\end{eqnarray*}
where the first equality follows from (\ref{pf:e}), and the second equality follows from the relation $(\Sigma_{1}^{(i)})^{-1}=\Sigma^{-1}-\Sigma^{-1}(\Sigma_{1}^{(i)}-\Sigma)(\Sigma_{1}^{(i)})^{-1}$.
By the same argument as in (\ref{pf:M23}), it holds $\hat{\sigma}_{132,\text{HC3}}^{2}=O_p\left(\mathcal{E}_{2}\kappa^{3/2}(\log p)^{1/2}\right)=o_{p}(\mathcal{E}_{2}\kappa)$.
For $\hat{\sigma}_{131,\text{HC3}}^{2}$, we have 
\begin{eqnarray*}
 & & {\mathbb E}[(\hat{\sigma}_{131,\text{HC3}}^{2})^2] \\
 & = & {\mathbb E}\left[\left(\frac{2n}{n_{1}}\left\{ \frac{1}{n}\sum_{i=1}^{n}\left(\frac{T_{i}}{\pi}-1\right)z_{i}e_{i}(1)\right\} ^{\prime}\Sigma^{-1}\left\{ \frac{1}{n}\sum_{j\neq i}^{n}\left(\frac{T_{j}}{\pi}-1\right)z_{j}e_{j}(1)-\frac{1}{n}z_{i}e_{i}(1)\right\}\right)^2\right]\\
& \lesssim & {\mathbb E}\left[\frac{1}{n^2}\sum_{\substack{i,i_1,j,j_1=1\\ j\neq i\\j_1\neq i}}^n\left(\frac{T_{i}}{\pi}-1\right)\left(\frac{T_{i_1}}{\pi}-1\right)\left(\frac{T_{j}}{\pi}-1\right)\left(\frac{T_{j_1}}{\pi}-1\right)P_{ij}P_{i_1j_1}e_{i}(1)e_{i_1}(1)e_{j}(1)e_{j_1}(1)\right]\\
& & +{\mathbb E}\left[\frac{1}{n^2}\sum_{i,i_1=1}^n\left(\frac{T_{i}}{\pi}-1\right)\left(\frac{T_{i_1}}{\pi}-1\right)P_{ii}P_{i_1i_1}e_i(1)^2e_{i_1}(1)^2\right]\\
& \lesssim & \frac{1}{n^2}\sum_{i,j} P_{ij}^2 e_i(1)^2e_j(1)^2+\frac{1}{n^2}\sum_{i=1}^n P_{ii}^2e_i(1)^4\\
&\le & {\mathcal E}_2^2\kappa^4+\frac{{\mathcal E}_{\infty}^2}{n}{\mathcal E}_2\kappa^2 =o(({\mathcal E}_2\kappa)^2),
\end{eqnarray*}
where the third inequality follows from $P_{ij}^2 \le P_{ii}^2P_{jj}^2 \le \kappa^4$.
Thus, $\hat{\sigma}_{131,\text{HC3}}^{2}=o_p({\mathcal E}_2\kappa)$.

For $\hat{\sigma}_{12,\text{HC3}}^2$, repeated application of the relation $(\Sigma_{1}^{(i)})^{-1}=\Sigma^{-1}-\Sigma^{-1}(\Sigma_{1}^{(i)}-\Sigma)(\Sigma_{1}^{(i)})^{-1}$ yields
\begin{align*}
\hat{\sigma}_{12,\text{HC3}}^2
&=\frac{1}{\pi}\left(\frac{1}{n}\sum_{j\neq i}^{n}\frac{T_{j}}{\pi}z_{j}e_{j}(1)\right)^{\prime}\Sigma^{-1}\Sigma_1\Sigma^{-1}\left(\frac{1}{n}\sum_{j\neq i}^{n}\frac{T_{j}}{\pi}z_{j}e_{j}(1)\right)+o_p({\mathcal E}_2 \kappa) \\
&=\frac{1}{\pi}\left(\frac{1}{n}\sum_{j\neq i}^{n}\frac{T_{j}}{\pi}z_{j}e_{j}(1)\right)^{\prime}\Sigma^{-1}\left(\frac{1}{n}\sum_{j\neq i}^{n}\frac{T_{j}}{\pi}z_{j}e_{j}(1)\right)+o_p({\mathcal E}_2 \kappa)\\
&=\hat{\sigma}_{121,\text{HC0}}^2+o_p({\mathcal E}_2 \kappa),
\end{align*}
where the third equality follows from the relation $\frac{1}{n}\sum_{j\neq i}^{n}\frac{T_{j}}{\pi}z_{j}e_{j}(1)=\frac{1}{n}\sum_{j=1}^{n}\left(\frac{T_{j}}{\pi}-1\right)z_{j}e_{j}(1)-\frac{1}{n}\frac{T_i}{\pi}z_{i}e_{i}(1)$ and $\left\|\frac{1}{n}\frac{T_i}{\pi}z_{i}e_{i}(1)\right\|=O_p\left(n^{-1/2}({\mathcal E}_{\infty}\kappa\right)^{1/2})=o_p\left(\sqrt{{\mathcal E}_2\kappa}\right)$.

Combining these results, we obtain
\begin{equation}\label{pf:HC31}
\hat{\sigma}_{1,\text{HC3}}^{2}=\frac{n}{n_1(n-1)}\sum_{i=1}^n e_i(1)^2+\frac{n_0}{n_1^2}\sum_{i=1}^n P_{ii}e_i(1)^2+o_p({\mathcal E}_2\kappa).
\end{equation}
The same argument yields $\hat{\sigma}_{0,\text{HC3}}^{2}=\frac{n}{n_{0}(n-1)}\sum_{i=1}^{n}e_{i}(0)^{2}+\frac{n_{1}}{n_{0}^{2}}\sum_{i=1}^{n}P_{ii}e_{i}(0)^{2}+o_p({\mathcal E}_2\kappa)$, and the conclusion for $\hat{\sigma}_{\text{HC3}}^{2}$ follows.

\subsection{Proof of Corollary \ref{cor:mHC3}}

We first note that
\begin{align*}
&\left\{\frac{n}{n_{1}(n-1)}\sum_{i=1}^{n}e_{i}(1)^{2}+\frac{n}{n_{0}(n-1)}\sum_{i=1}^{n}e_{i}(0)^{2}+\frac{n_{0}}{n_{1}^{2}}\sum_{i=1}^{n}P_{ii}e_{i}(1)^{2}+\frac{n_{1}}{n_{0}^{2}}\sum_{i=1}^{n}P_{ii}e_{i}(0)^{2}\right\}\\
&+  \frac{n_1-n_0}{n_1 n}\sum_{i=1}^n P_{ii}e_i(1)^2 + \frac{n_0-n_1}{n_0 n}\sum_{i=1}^n P_{ii}e_i(0)^2\\
&+ \frac{n_{0}^{2}}{n_{1}^{2} n}\sum_{i=1}^{n}\sum_{j\neq i}^{n}P_{ij}^{2}e_{i}(1)e_{j}(1)+\frac{n_{1}^{2}}{n_{0}^{2}n}\sum_{i=1}^{n}\sum_{j\neq i}^{n}P_{ij}^{2}e_{i}(0)e_{j}(0) 
 -\frac{2}{n}\sum_{i=1}^{n}\sum_{j\neq i}^{n}P_{ij}^{2}e_{i}(1)e_{j}(0)\\
 &=\sigma_L^{2}+\sigma_W^{2}+\frac{1}{n-1}\sum_{i=1}^n (e_i(1)-e_i(0))^2+\frac{1}{n}\sum_{i=1}^nP_{ii}(e_i(1)-e_i(0))^2\\
 &\ge \sigma_L^{2}+\sigma_W^{2}.
\end{align*}
Hence, combining with the result of $\hat{\sigma}_{\text{HC3}}^2$ in Theorem \ref{thm:var1}, it is enough to show that
\begin{align}
\frac{n_1-n_0}{n_1^2}\sum_{i=1}^n T_iP_{ii}\tilde{e}_i^2&=\frac{n_1-n_0}{n_1 n}\sum_{i=1}^n P_{ii}e_i(1)^2(1+o_p(1)), \label{eq:cor1}\\
 \frac{n_0-n_1}{n_0^2}\sum_{i=1}^n (1-T_i)P_{ii}\tilde{e}_i^2&=\frac{n_0-n_1}{n_0 n}\sum_{i=1}^n P_{ii}e_i(0)^2(1+o_p(1)),\label{eq:cor2}\\
\frac{n_{0}^{2}n}{n_{1}^{4}}\sum_{i=1}^{n}\sum_{j\neq i}^{n}P_{ij}^{2}T_{i}T_{j}\tilde{e}_{i}\tilde{e}_{j}&=\frac{n_{0}^{2}}{n_{1}^{2}n}\sum_{i=1}^{n}\sum_{j\neq i}^{n}P_{ij}^{2}e_{i}(1)e_{j}(1)(1+o_p(1)),\label{eq:cor3}\\
\frac{n_{1}^{2}n}{n_{0}^{4}}\sum_{i=1}^{n}\sum_{j\neq i}^{n}P_{ij}^{2}(1-T_{i})(1-T_{j})\tilde{e}_{i}\tilde{e}_{j}&= \frac{n_{1}^{2}}{n_{0}^{2}n}\sum_{i=1}^{n}\sum_{j\neq i}^{n}P_{ij}^{2}e_{i}(0)e_{j}(0)(1+o_p(1)),\label{eq:cor4}\\
-\frac{2n}{n_{0}n_{1}}\sum_{i=1}^{n}\sum_{j\neq i}^{n}P_{ij}^{2}T_{i}(1-T_{j})\tilde{e}_{i}\tilde{e}_{j}&=-\frac{2}{n}\sum_{i=1}^{n}\sum_{j\neq i}^{n}P_{ij}^{2}e_{i}(1)e_{j}(0)(1+o_p(1)).\label{eq:cor5}
\end{align}
For (\ref{eq:cor1}), it holds
\begin{align*}
&\frac{n_1-n_0}{n_1^2}\sum_{i=1}^n T_iP_{ii}\tilde{e}_i^2
=\frac{n_1-n_0}{n_1n}\sum_{i=1}^n \frac{T_i}{\pi}P_{ii} \left\{e_i(1)-z_i^{\prime}(\hat{\beta}_1^{(i)}-\beta_1)\right\}^2\\
&=\frac{n_1-n_0}{n_1n}\sum_{i=1}^n \frac{T_i}{\pi}P_{ii}e_i(1)^2
+\frac{n_1-n_0}{n_1n}\sum_{i=1}^n \left(\frac{1}{n}\sum_{j\neq i}^n \frac{T_j}{\pi}z_je_j(1)\right)^{\prime}(\Sigma_1^{(i)})^{-1}\frac{T_i}{\pi}P_{ii}z_iz_i^{\prime}(\Sigma_1^{(i)})^{-1}\left(\frac{1}{n}\sum_{j\neq i}^n \frac{T_j}{\pi}z_je_j(1)\right)\\
&-\frac{2(n_1-n_0)}{n_1n}\sum_{i=1}^n \frac{T_i}{\pi}P_{ii}e_i(1)z_i^{\prime}(\Sigma_1^{(i)})^{-1}\left(\frac{1}{n}\sum_{j\neq i}^n \frac{T_j}{\pi}z_je_j(1)\right)\\
&=\frac{n_1-n_0}{n_1 n}\sum_{i=1}^n \frac{T_i}{\pi}P_{ii}e_i(1)^2(1+o_p(1))\\
&=\frac{n_1-n_0}{n_1 n}\sum_{i=1}^n P_{ii}e_i(1)^2(1+o_p(1)),
\end{align*}
where the third equality follows from
\begin{align*}
&\left|\frac{n_1-n_0}{n_1n}\sum_{i=1}^n \left(\frac{1}{n}\sum_{j\neq i}^n \frac{T_j}{\pi}z_je_j(1)\right)^{\prime}(\Sigma_1^{(i)})^{-1}\frac{T_i}{\pi}P_{ii}z_iz_i^{\prime}(\Sigma_1^{(i)})^{-1}\left(\frac{1}{n}\sum_{j\neq i}^n \frac{T_j}{\pi}z_je_j(1)\right)\right|\\
& \lesssim \kappa\left|\frac{1}{n}\sum_{i=1}^n \left(\frac{1}{n}\sum_{j\neq i}^n \frac{T_j}{\pi}z_je_j(1)\right)^{\prime}(\Sigma_1^{(i)})^{-1}\frac{T_i}{\pi}z_iz_i^{\prime}(\Sigma_1^{(i)})^{-1}\left(\frac{1}{n}\sum_{j\neq i}^n \frac{T_j}{\pi}z_je_j(1)\right)\right|\\
&\sim \kappa \hat{\sigma}_{12, \text{HC3}}^2=o_p({\mathcal E}_2 \kappa),
\end{align*}
and
\begin{align*}
&\left|\frac{2(n_1-n_0)}{n_1n}\sum_{i=1}^n \frac{T_i}{\pi}P_{ii}e_i(1)z_i^{\prime}(\Sigma_1^{(i)})^{-1}\left(\frac{1}{n}\sum_{j\neq i}^n \frac{T_j}{\pi}z_je_j(1)\right)\right|\\
&\lesssim \sqrt{\frac{1}{n}\sum_{i=1}^n P_{ii}^2e_i(1)^2}\sqrt{\frac{1}{n}\sum_{i=1}^n\left(\frac{1}{n}\sum_{j\neq i}^n \frac{T_j}{\pi}z_je_j(1)\right)^{\prime}(\Sigma_1^{(i)})^{-1}z_iz_i^{\prime}(\Sigma_1^{(i)})^{-1}\left(\frac{1}{n}\sum_{j\neq i}^n \frac{T_j}{\pi}z_je_j(1)\right)}\\
&\sim  \sqrt{\frac{1}{n}\sum_{i=1}^n P_{ii}^2e_i(1)^2}\sqrt{\hat{\sigma}_{12, \text{HC3}}^2}
=o_p({\mathcal E}_2 \kappa),
\end{align*}
and the fourth equality follows from the law of large numbers.
(\ref{eq:cor2}) -- (\ref{eq:cor5}) follow from the same argument. Thus the conclusion follows.
}

\subsection{Auxiliary Lemmas}
\begin{lem}
\label{lem:T} For random variables $\{T_{i}\}_{i=1}^{n}$ sampled
without replacement with probability $\mathbb{P}\{T_{i}=1\}=n_{1}/n=\pi\to\pi_{\infty}\in(0,1)$,
it holds 
\begin{align*}
 & \mathbb{E}[T_{i}]=\pi,\qquad\mathbb{V}(T_{i})=\pi(1-\pi),\qquad\mathbb{E}[(T_{i}-\pi)(T_{j}-\pi)]=-\frac{\pi(1-\pi)}{n-1},\\
 & \mathbb{E}[(T_{i}-\pi)^{3}]=\pi(1-\pi)(1-2\pi),\\
 & \mathbb{E}[(T_{i}-\pi)^{2}(T_{j}-\pi)]=-\frac{\pi(1-\pi)(1-2\pi)}{n-1},\\
 & \mathbb{E}[(T_{i}-\pi)(T_{j}-\pi)(T_{k}-\pi)]=\frac{2\pi(1-\pi)(1-2\pi)}{(n-1)(n-2)},\\
 & \mathbb{E}[(T_{i}-\pi)^{4}]=\pi(1-\pi)\{1-3\pi(1-\pi)\}\\
 & \mathbb{E}[(T_{i}-\pi)^{2}(T_{j}-\pi)^{2}]=\frac{n}{n-1}\pi^{2}(1-\pi)^{2}+O(n^{-1}),\\
 & \mathbb{E}[(T_{i}-\pi)^{2}(T_{j}-\pi)(T_{k}-\pi)]=-\frac{n}{(n-1)(n-2)}\pi^{2}(1-\pi)^{2}+O(n^{-2}),\\
 & \mathbb{E}[(T_{i}-\pi)(T_{j}-\pi)(T_{k}-\pi)(T_{l}-\pi)]=\frac{3n}{n-1}\frac{\pi^{2}(1-\pi)^{2}}{(n-2)(n-3)}+O(n^{-3}),\\
 & \mathbb{E}[(T_{i}-\pi)^{3}(T_{j}-\pi)^{3}]=O(1),\\
 & \mathbb{E}[(T_{i}-\pi)^{4}(T_{j}-\pi)^{2}]=O(1),\\
 & \mathbb{E}[(T_{i}-\pi)^{5}(T_{j}-\pi)]=O(n^{-1}),\\
 & \mathbb{E}[(T_{q}-\pi)^{4}(T_{i}-\pi)(T_{j}-\pi)]=O(n^{-1}),\\
 & \mathbb{E}[(T_{q}-\pi)^{2}(T_{i}-\pi)^{2}(T_{j}-\pi)(T_{k}-\pi)]=O(n^{-1}),\\
 & \mathbb{E}[(T_{q}-\pi)^{3}(T_{i}-\pi)(T_{j}-\pi)(T_{k}-\pi)]=O(n^{-2}),\\
 & \mathbb{E}[(T_{q}-\pi)^{2}(T_{i}-\pi)(T_{j}-\pi)(T_{k}-\pi)(T_{l}-\pi)]=O(n^{-2}),\\
 & \mathbb{E}[(T_{i}-\pi)(T_{j}-\pi)(T_{k}-\pi)(T_{l}-\pi)(T_{m}-\pi)(T_{q}-\pi)]=O(n^{-3}).
\end{align*}
for any six mutually distinctive $i,j,k,l,m,q\in\{1,...,n\}$. 
\end{lem}

This Lemma is a direct implication of the following Lemmas \ref{lem:moments_1},
\ref{lem:moments_2}, \ref{lem:moments_3}, \ref{lem:moments_4},
and \ref{lem:moments_5}. 
\begin{lem}
\label{lem:moments_1} Under the conditions of Lemma \ref{lem:T},
it holds 
\begin{align*}
 & \mathbb{E}[T_{i}T_{j}]=\pi^{2}+O(n^{-1}),\qquad\mathbb{E}[T_{i}T_{j}T_{k}]=\pi^{3}+O(n^{-1}),\\
 & \mathbb{E}[T_{i}T_{j}T_{k}T_{l}]=\pi^{4}+O(n^{-1}),\qquad\mathbb{E}[T_{i}T_{j}T_{k}T_{l}T_{m}]=\pi^{5}+O(n^{-1}),\\
 & \mathbb{E}[T_{i}T_{j}T_{k}T_{l}T_{m}T_{q}]=\pi^{6}+O(n^{-1}).
\end{align*}
\begin{proof}
Observe that 
\begin{align*}
\mathbb{E}[T_{j}T_{k}]= & \frac{n_{1}(n_{1}-1)}{n(n-1)}=\pi^{2}+\frac{n_{1}(n_{1}-n)}{n^{2}(n-1)}=\pi^{2}+O(n^{-1}),\\
\mathbb{E}[T_{j}T_{k}T_{l}]= & \frac{n_{1}(n_{1}-1)(n_{1}-2)}{n(n-1)(n-2)}=\pi^{3}+\frac{n_{1}[n_{1}^{2}(3n-2)-n^{2}(3n_{1}-2)]}{n^{3}(n-1)(n-2)}=\pi^{3}+O(n^{-1}).
\end{align*}
Similarly, direct calculations yield that 
\begin{align*}
\mathbb{E}[T_{i}T_{j}T_{k}T_{l}]= & \pi^{4}+\frac{n_{1}[n_{1}^{3}(6n^{2}-11n+6)-n^{3}(6n_{1}^{2}-11n_{1}+6)]}{n^{4}(n-1)(n-2)(n-3)}=\pi^{4}+O(n^{-1}),\\
\mathbb{E}[T_{q}T_{i}T_{j}T_{k}T_{l}]= & \pi^{5}+\frac{n_{1}[n_{1}^{4}(10n^{3}-35n^{2}+50n-24)-n^{4}(10n_{1}^{3}-35n_{1}^{2}+50n_{1}-24)]}{n^{5}(n-1)(n-2)(n-3)(n-4)}.
\end{align*}
Finally, it holds that 
\begin{align*}
 & \mathbb{E}[T_{i}T_{j}T_{k}T_{l}T_{m}T_{q}]\\
= & \pi^{6}+\frac{n_{1}[n_{1}^{5}(15n^{4}-85n^{3}+225n^{2}-274n+120)-n^{5}(15n_{1}^{4}-85n_{1}^{3}+225n_{1}^{2}-274n_{1}+120)]}{n^{6}(n-1)(n-2)(n-3)(n-4)(n-5)}\\
= & \pi^{6}+O(n^{-1}).
\end{align*}
\end{proof}
\end{lem}

\begin{lem}
\label{lem:moments_2} Under the conditions of Lemma \ref{lem:T},
it holds 
\begin{align*}
 & \mathbb{E}[(T_{j}-\pi)^{2}(T_{k}-\pi)^{2}]=\pi^{2}(1-\pi)^{2}+O(n^{-1}),\\
 & \mathbb{E}[(T_{j}-\pi)^{2}(T_{k}-\pi)^{2}(T_{l}-\pi)^{2}]=-\pi^{3}(1-\pi)^{3}+O(n^{-1}),\\
 & \mathbb{E}[(T_{j}-\pi)^{4}]=\pi(1-\pi)[1-3\pi(1-\pi)].
\end{align*}

\begin{proof}
For the first result, by the fact that $(T_{j}-\pi)^{2}=(1-2\pi)T_{j}+\pi^{2}$,
we have 
\begin{align*}
\mathbb{E}[(T_{j}-\pi)^{2}(T_{k}-\pi)^{2}]= & \mathbb{E}[(T_{j}(1-2\pi)+\pi^{2})(T_{k}(1-2\pi)+\pi^{2})]\\
= & \frac{n_{1}(n_{1}-1)}{n(n-1)}\left(1-2\frac{n_{1}}{n}\right)^{2}+\frac{2n_{1}^{3}}{n^{3}}\left(1-2\frac{n_{1}}{n}\right)+\frac{n_{1}^{4}}{n^{4}}\\
= & \frac{n}{n-1}\pi^{2}(1-\pi)^{2}+O(n^{-1}).
\end{align*}

For the second result, using the fact that $(T_{j}-\pi)^{2}=(1-2\pi)T_{j}+\pi^{2}$,
one has 
\begin{align*}
 & \mathbb{E}[(T_{j}-\pi)^{2}(T_{k}-\pi)^{2}(T_{l}-\pi)^{2}]\\
= & (1-2\pi)^{3}\mathbb{E}[T_{j}T_{k}T_{l}]+(1-2\pi)^{2}\pi^{2}3\mathbb{E}[T_{j}T_{k}]+(1-2\pi)\pi^{4}3\mathbb{E}[T_{j}]+\pi^{6}\\
= & -\pi^{3}(1-\pi)^{3}+O(n^{-1}).
\end{align*}

Finally, notice that 
\begin{align*}
\mathbb{E}[(T_{i}-\pi)^{4}]= & \mathbb{E}[((1-2\pi)T_{i}+\pi^{2})^{2}]=-3\pi^{4}+6\pi^{3}-4\pi^{2}+\pi=\pi(1-\pi)[1-3\pi(1-\pi)].
\end{align*}
\end{proof}
\end{lem}


\begin{lem}
\label{lem:moments_3} Under the conditions of Lemma \ref{lem:T},
it holds 
\begin{align*}
 & \mathbb{E}[(T_{i}-\pi)(T_{j}-\pi)]=-\frac{\pi(1-\pi)}{(n-1)}=O(n^{-1}),\\
 & \mathbb{E}[(T_{i}-\pi)(T_{j}-\pi)(T_{k}-\pi)]=\frac{2\pi(1-\pi)(1-2\pi)}{(n-1)(n-2)}=O(n^{-2}),\\
 & \mathbb{E}[(T_{i}-\pi)(T_{j}-\pi)(T_{k}-\pi)(T_{l}-\pi)]=\frac{3n\pi^{2}(1-\pi)}{(n-1)(n-2)(n-3)}+O(n^{-3})=O(n^{-2}).
\end{align*}
\begin{proof}
First, observe that 
\begin{align*}
\mathbb{E}[(T_{i}-\pi)(T_{j}-\pi)]= & \mathbb{E}[T_{i}T_{j}]-\pi^{2}=\frac{n_{1}(n_{1}-n)}{n^{2}(n-1)}=-\frac{\pi(1-\pi)}{(n-1)}=O(n^{-1}).
\end{align*}
and, similarly 
\begin{align*}
\mathbb{E}[(T_{i}-\pi)(T_{j}-\pi)(T_{k}-\pi)]= & \frac{n_{1}(n_{1}-1)(n_{1}-2)}{n(n-1)(n-2)}-3\frac{n_{1}^{2}(n_{1}-1)}{n^{2}(n-1)}+2\frac{n_{1}^{3}}{n^{3}}\\
= & \frac{2\pi(1-\pi)(1-2\pi)}{(n-1)(n-2)}=O(n^{-2}),
\end{align*}
which shows the first statement.

Secondly, note that using Lemma 1, direct calculations yield that
\begin{align*}
\mathbb{E}[(T_{i}-\pi)(T_{j}-\pi)(T_{k}-\pi)(T_{l}-\pi)]= & \mathbb{E}[T_{i}T_{j}T_{k}T_{l}]-4\pi\mathbb{E}[T_{i}T_{j}T_{k}]+6\pi^{3}\mathbb{E}[T_{i}T_{j}]-3\pi^{4}\\
= & \frac{3n_{1}(n_{1}-n)(n_{1}^{2}n-n_{1}n^{2}+6n_{1}(n_{1}-n)+2n^{2})}{n^{4}(n-1)(n-2)(n-3)}\\
= & \frac{3n\pi^{2}(1-\pi)}{(n-1)(n-2)(n-3)}+O(n^{-3})=O(n^{-2}).
\end{align*}
\end{proof}
\end{lem}


\begin{lem}
\label{lem:moments_4} Under the conditions of Lemma \ref{lem:T},
it holds 
\begin{align*}
 & \mathbb{E}[(T_{i}-\pi)^{2}(T_{j}-\pi)]=-\frac{\pi(1-\pi)(1-2\pi)}{(n-1)},\quad\mathbb{E}[(T_{i}-\pi)^{2}(T_{j}-\pi)(T_{k}-\pi)]=-\frac{n\pi^{2}(1-\pi)^{2}}{(n-1)(n-2)}+O(n^{-2}).\\
\end{align*}
\begin{proof}
The first result follows from the calculation that 
\begin{align*}
\mathbb{E}[(T_{i}-\pi)^{2}(T_{j}-\pi)]= & \mathbb{E}[((1-2\pi)T_{i}+\pi^{2})(T_{j}-\pi)]=-\frac{\pi(1-\pi)(1-2\pi)}{(n-1)}.
\end{align*}

For the second statement, note that 
\begin{align*}
\mathbb{E}[(T_{i}-\pi)^{2}(T_{j}-\pi)(T_{k}-\pi)]= & \frac{n_{1}n_{0}(n_{1}^{2}n-n_{1}n^{2})}{n^{4}(n-1)(n-2)}+\frac{n_{1}n_{0}(6n_{1}^{2}-6n_{1}n+2n^{2})}{n^{4}(n-1)(n-2)}\\
= & -\frac{n\pi^{2}(1-\pi)^{2}}{(n-1)(n-2)}+O(n^{-2}).
\end{align*}
\end{proof}
\end{lem}


\begin{lem}
\label{lem:moments_5} Under the conditions of Lemma \ref{lem:T},
it holds 
\begin{align*}
 & \mathbb{E}[(T_{i}-\pi)(T_{j}-\pi)(T_{k}-\pi)(T_{l}-\pi)(T_{m}-\pi)(T_{q}-\pi)]=O(n^{-3}),\\
 & \mathbb{E}[(T_{i}-\pi)^{3}(T_{j}-\pi)^{3}]=O(1),\qquad\mathbb{E}[(T_{i}-\pi)^{4}(T_{j}-\pi)^{2}]=O(1),\\
 & \mathbb{E}[(T_{i}-\pi)^{5}(T_{j}-\pi)]=O(n^{-1}),\qquad\mathbb{E}[(T_{q}-\pi)^{4}(T_{i}-\pi)(T_{j}-\pi)]=O(n^{-1}),\\
 & \mathbb{E}[(T_{q}-\pi)^{2}(T_{i}-\pi)^{2}(T_{j}-\pi)(T_{k}-\pi)]=O(n^{-1}),\\
 & \mathbb{E}[(T_{q}-\pi)^{3}(T_{i}-\pi)(T_{j}-\pi)(T_{k}-\pi)]=O(n^{-2}),\\
 & \mathbb{E}[(T_{q}-\pi)^{2}(T_{i}-\pi)(T_{j}-\pi)(T_{k}-\pi)(T_{l}-\pi)]=O(n^{-2}).
\end{align*}

\begin{proof}
By a brute force calculation, we have 
\begin{align*}
 & \mathbb{E}[(T_{q}-\pi)(T_{i}-\pi)(T_{j}-\pi)(T_{k}-\pi)(T_{l}-\pi)(T_{m}-\pi)]\\
= & \mathbb{E}[T_{q}T_{i}T_{j}T_{k}T_{l}T_{m}]-6\pi\mathbb{E}[T_{q}T_{i}T_{j}T_{k}T_{l}]+15\pi^{2}\mathbb{E}[T_{q}T_{i}T_{j}T_{k}]-20\pi^{3}\mathbb{E}[T_{q}T_{i}T_{j}]+15\pi^{4}\mathbb{E}[T_{q}T_{i}]-5\pi^{6}\\
= & \frac{5n_{1}[-24n^{5}+2n_{1}n^{4}(72+13n)-3n_{1}^{2}n^{3}(120+46n+n^{2})+n_{1}^{5}(120+86n+3n^{2})]}{n^{6}(n-1)(n-2)(n-3)(n-4)(n-5)}\\
 & +\frac{5n_{1}[-3n_{1}^{4}n(120+86n+3n^{2})+n_{1}^{3}n^{2}(480+284n+9n^{2})]}{n^{6}(n-1)(n-2)(n-3)(n-4)(n-5)}\\
= & O(n^{8}n^{-11})=O(n^{-3}).
\end{align*}
Next, note that 
\begin{align*}
(T_{i}-\pi)^{3}=(3\pi^{2}-3\pi+1)T_{i}-\pi^{3},
\end{align*}
Thus 
\begin{align*}
\mathbb{E}[(T_{i}-\pi)^{3}(T_{j}-\pi)^{3}]= & \mathbb{E}[((3\pi^{2}-3\pi+1)T_{i}-\pi^{3})((3\pi^{2}-3\pi+1)T_{j}-\pi^{3})]\\
= & \pi-6\pi^{2}+15\pi^{3}-20\pi^{4}+15\pi^{5}-7\pi^{6}.
\end{align*}
The third statement can be shown similarly. Next, as 
\begin{align*}
(T_{i}-\pi)^{5}=(5\pi^{4}-10\pi^{3}+10\pi^{2}-5\pi+1)T_{i}-\pi^{5},
\end{align*}
we have 
\begin{align*}
\mathbb{E}[(T_{i}-\pi)^{5}(T_{j}-\pi)]= & (5\pi^{4}-10\pi^{3}+10\pi^{2}-5\pi+1)\mathbb{E}[T_{i}T_{j}]-(5\pi^{4}-10\pi^{3}+10\pi^{2}-5\pi+1)\pi^{2}-\pi^{6}+\pi^{6}\\
= & O(n^{-1}).
\end{align*}
Next, by Lemma \ref{lem:moments_2}, it holds that 
\begin{align*}
 & \mathbb{E}[(T_{q}-\pi)^{4}(T_{i}-\pi)(T_{j}-\pi)]=O(1)\mathbb{E}[(T_{i}-\pi)(T_{j}-\pi)]=O(n^{-1}),\\
 & \mathbb{E}[(T_{q}-\pi)^{2}(T_{i}-\pi)^{2}(T_{j}-\pi)(T_{k}-\pi)]=O(1)\mathbb{E}[(T_{j}-\pi)(T_{k}-\pi)]=O(n^{-1}),\\
 & \mathbb{E}[(T_{q}-\pi)^{3}(T_{i}-\pi)(T_{j}-\pi)(T_{k}-\pi)]=O(1)\mathbb{E}[(T_{i}-\pi)(T_{j}-\pi)(T_{k}-\pi)]=O(n^{-2}).
\end{align*}
Similarly, by Lemma \ref{lem:moments_3}, we have 
\begin{align*}
 & \mathbb{E}[(T_{q}-\pi)^{2}(T_{i}-\pi)(T_{j}-\pi)(T_{k}-\pi)(T_{l}-\pi)]=O(1)\mathbb{E}[(T_{i}-\pi)(T_{j}-\pi)(T_{k}-\pi)(T_{l}-\pi)]=O(n^{-2}),\\
 & \mathbb{E}[(T_{q}-\pi)^{2}(T_{i}-\pi)(T_{j}-\pi)(T_{k}-\pi)(T_{l}-\pi)]=O(1)\mathbb{E}[(T_{i}-\pi)(T_{j}-\pi)(T_{k}-\pi)(T_{l}-\pi)]=O(n^{-2}).
\end{align*}
\end{proof}
\end{lem}

\newpage{}

\qquad{}\\
  \bibliographystyle{econ-econometrica}
\bibliography{RCTbib}

\end{document}